**Functional neuroanatomy of meditation: A review and meta-analysis of 78 functional neuroimaging investigations**

Kieran C. R. Fox[a,*], Matthew L. Dixon[a], Savannah Nijeboer[a], Manesh Girn[a], James L. Floman[b], Michael Lifshitz[c], Melissa Ellamil[d], Peter Sedlmeier[e], and Kalina Christoff[a,f]

[a] Department of Psychology, University of British Columbia, 2136 West Mall, Vancouver, B.C., V6T 1Z4 Canada

[b] Department of Educational and Counselling Psychology, and Special Education, University of British Columbia, 2125 Main Mall, Vancouver, B.C., V6T 1Z4

[c] Integrated Program in Neuroscience, McGill University, 3775 University St., Montreal, QC, H3A 2B4

[d] Neuroanatomy and Connectivity Research Group, Max Planck Institute for Human Cognitive and Brain Sciences, Stephanstrasse 1a, Leipzig, Germany 04103.

[e] Institut für Psychologie, Technische Universität Chemnitz, 43 Wilhelm-Raabe Street, Chemnitz, Germany

[f] Brain Research Centre, University of British Columbia, 2211 Wesbrook Mall, Vancouver, B.C., V6T 2B5 Canada

[*] To whom correspondence may be addressed (at address [a] above). Telephone: 1-778-968-3334; Fax: 1-604-822-6923; E-mail: kfox@psych.ubc.ca

## Abstract

Meditation is a family of mental practices that encompasses a wide array of techniques employing distinctive mental strategies. We systematically reviewed 78 functional neuroimaging (fMRI and PET) studies of meditation, and used activation likelihood estimation to meta-analyze 257 peak foci from 31 experiments involving 527 participants. We found reliably dissociable patterns of brain activation and deactivation for four common styles of meditation (focused attention, mantra recitation, open monitoring, and compassion/loving-kindness), and suggestive differences for three others (visualization, sense-withdrawal, and non-dual awareness practices). Overall, dissociable activation patterns are congruent with the psychological and behavioral aims of each practice. Some brain areas are recruited consistently across multiple techniques – including insula, pre/supplementary motor cortices, dorsal anterior cingulate cortex, and frontopolar cortex – but convergence is the exception rather than the rule. A preliminary effect-size meta-analysis found medium effects for both activations (d = .59) and deactivations (d = -.74), suggesting potential practical significance. Our meta-analysis supports the neurophysiological dissociability of meditation practices, but also raises many methodological concerns and suggests avenues for future research.

**Keywords:** meditation; open monitoring; focused attention; functional magnetic resonance imaging; meta-analysis; activation likelihood estimation

## 1. Introduction

Meditation has been used as a tool to train the mind for thousands of years (Anālayo, 2003, Iyengar, 2005). Broadly speaking, meditation practices involve the monitoring and regulation of attention and emotion (Lutz et al., 2008b, Tang et al., 2015). Meditation can be either directed outward to particular objects and sensory stimuli or turned inward to the workings of the mind and felt experiences of the body. A common principle across all forms of meditation is that through specific and regular practice, the ability to monitor and regulate mental-physical processes can be progressively developed – the way performance on any other cognitive or motor skill can be improved with practice (Ericsson et al., 1993, Slagter et al., 2011, Vago, 2014). Interest in the neural basis of meditative practices has increased enormously over the past decade, and mounting empirical evidence suggests that meditation indeed leads to or is associated with significant changes in cognitive and affective processing (Sedlmeier et al., 2012), as well as alterations in brain structure (Fox et al., 2014) and function (Cahn and Polich, 2006).

Although meditative practices now garner serious interest from the cognitive neuroscience community (Fig. 1), meditation still tends to be viewed as a somewhat uniform practice (e.g., (Sperduti et al., 2012). The umbrella-term 'meditation' encompasses a wide variety of distinct practices with specific goals and methods, however. The varying methods and scope of particular forms of meditation include focalizing and sustaining attention (Wallace, 2006), generating and maintaining complex visual imagery (Kozhevnikov et al., 2009), improving emotion regulation and well-being (Chambers et al., 2009, Sedlmeier et al., 2012), and deepening compassion for others (Galante et al., 2014).

Despite the diversity of meditation practices, few behavioral, clinical, or neuroimaging studies have directly compared meditation styles within or across studies. As such, little is known about whether functional neural differences underlie each type of practice (although see (Lou et al., 1999, Manna et al., 2010, Brewer et al., 2011, Lee et al., 2012, Tomasino et al., 2013). Although mounting empirical evidence suggests that these distinctive psychological practices may be dissociable at the neurophysiological level (Lutz et al., 2008b, Travis and Shear, 2010a, Josipovic et al., 2011, Tomasino et al., 2013), no comprehensive overview has adequately addressed this complex issue. We therefore attempted to synthesize the large and rapidly expanding (Fig. 1) body of work on the functional neuroanatomy of meditation practices in the present review and meta-analysis.

Our central hypothesis was a simple one: meditation practices distinct at the psychological level ($\Psi$) may be accompanied by dissociable activation patterns at the neurophysiological level ($\Phi$). Such a model describes a 'one-to-many' isomorphism between mind and brain: a particular psychological state or process is expected to have many neurophysiological correlates from which, ideally, a consistent pattern can be discerned (Cacioppo and Tassinary, 1990).

Supporting or disconfirming this prediction through empirical synthesis is challenging, however, because at present there are few (if any) 'objective' third-person measures that can confirm engagement in a particular subjective state or mental practice (Lutz and Thompson, 2003, Fazelpour and Thompson, 2015). Current investigations therefore necessarily rely on first-person reports that a specific mental practice is indeed being carried out as instructed or intended. Acknowledging this difficulty, we reasoned that if meditation practitioners are really engaging in the practices they report to be – and if these practices engender their intended effects and mental states – then distinctive patterns of brain activation and deactivation should accompany each technique.

In line with our hypothesis, it has been argued that various categories of meditation are characterized by differential electroencephalography (EEG) signatures (Travis and Shear, 2010a), cognitive-emotional effects (Sedlmeier et al., 2012), and potentially neuroanatomical alterations (Fox et al., 2014). In the present synthesis, we sought to build on prior reviews and meta-analyses by examining the possibility that comparable dissociations might exist with modalities that measure cerebral blood flow (positron emission tomography; PET) or blood-oxygenation level (functional magnetic resonance imaging; fMRI). Other modalities, such as EEG, have extremely high temporal resolution but very poor spatial resolution (or vice versa, in morphometric neuroimaging). Functional neuroimaging methods, on the other hand, are unique in their combination of good temporal and spatial resolution. This allows for precise localization of dynamic changes in brain activity, providing unified spatiotemporal neurophysiological data that could prove critical in differentiating one form of meditation practice from another.

Taken together, our central aims were to: (i) comprehensively review and meta-analyze the existing functional neuroimaging studies of meditation (using the meta-analytic method known as activation likelihood estimation, or ALE), and compare consistencies in brain activation and deactivation both within and across psychologically distinct meditation techniques; (ii) examine the magnitude of the effects that characterize these activation patterns, and address whether they suggest any practical significance; and (iii) articulate the various methodological challenges facing the emerging field of contemplative neuroscience (Caspi and Burleson, 2005, Thompson, 2009, Davidson, 2010, Davidson and Kaszniak, 2015), particularly with respect to functional neuroimaging studies of meditation.

*Figure 1.* The rapid increase in functional (fMRI and PET) and morphometric neuroimaging of meditation and related secular contemplative practices.

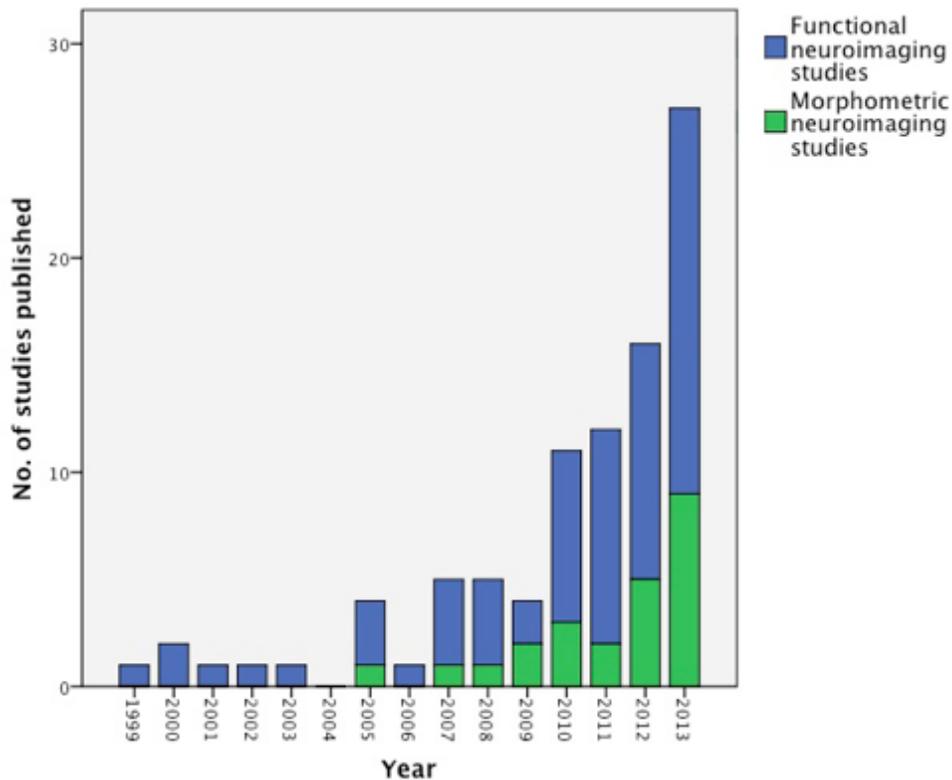

Functional neuroimaging (fMRI and PET) study counts based on the results of the present review; morphometric study counts based on our previous meta-analysis of this literature (Fox et al., 2014).

*1.1 Functional neuroimaging of meditation: The need for quantitative meta-analysis*

There are now close to a hundred functional neuroimaging studies of meditation (Fig. 1) – but can any conclusions be drawn from this large and fast growing literature? Determining reliable neural correlates of any mental process is a difficult task under the best of conditions. This is all the more true when the mental processes in question are subtle and largely subjective, and when the participants employed are highly experienced practitioners often operating within spiritual-religious frameworks. Despite the voluminous functional neuroimaging literature on meditation, we suspect that most researchers would be hard-pressed to provide a clear answer to the ostensibly simple question: What brain regions are reliably recruited, for instance, by focused attention meditation? We expect that similar challenges would face researchers seeking to describe other common categories of meditation. Part of the problem is that despite the rapid proliferation of neuroimaging studies of meditation in the past few years (Fig. 1), neither narrative (Rubia, 2009, Tang et al., 2015), nor meta-analytic (Sedlmeier et al., 2012, Fox et al., 2014), nor theoretical (Holzel et al., 2011, Vago and Silbersweig, 2012) synthesis of these results has been able to keep pace with the emergence of new and complex data. Given the numerous and often contradictory results reported to date, researchers interested in the neural correlates of meditation may find it difficult to decide among different findings. As experts on meta-analytic methodology recently framed the general problem of integrating findings:

> How should scientists proceed when results differ? First, it is clear how they should *not* proceed: they should not pretend that there is no problem, or decide that just one study (perhaps the most recent one, or the one they conducted, or a study chosen via some other equally arbitrary criterion) produced the correct finding. If results that are expected to be very similar show variability, the scientific instinct should be to account for the variability by further systematic work.  (Cooper and Hedges, 2009); p. 4)

It is in this spirit of systematic, quantitative synthesis that the present meta-analysis was conducted.

   Although the coordinate-based meta-analysis we employed here (see Methods) is far from the perfect (or only) method of achieving this kind of synthesis (Salimi-Khorshidi et al., 2009), it is a major improvement over narrative (and/or selective) surveys of the literature, which often lead to erroneous conclusions (Schmidt, 1992). We therefore echo other neuroscientists in arguing that quantitative meta-analysis of neuroimaging data is a potent (indeed necessary) tool for better understanding the neural basis of mental processes (Wager et al., 2007, Salimi-Khorshidi et al., 2009, Wager et al., 2009, Kober and Wager, 2010, Yarkoni et al., 2010, Hupé, 2015). More specifically, meta-analysis can provide: (i) a less-biased overview of an evidence base than a simple narrative review or qualitative survey of prior work (Schmidt, 1992); (ii) specific peaks of meta-analytic activation, rather than just broad regional results; (iii) statistically significant activation overlaps across studies, instead of merely indicating broad 'replications'; and (iv) a way to mitigate the lack of statistical power caused by small samples found in many neuroimaging studies – a problem especially prevalent in the study of meditation practitioners, where experienced individuals are difficult to recruit. For a deeper discussion of the limitations of these meta-analytic methods, however, see Section 5.11.

Moreover, persuasive arguments have also been made for a move toward a more 'cumulative' science that focuses on broad conclusions drawn from dozens or hundreds of independent investigations, as opposed to 'critical experiments' from isolated but influential studies (Schmidt, 1992, Yarkoni et al., 2010). Some statisticians have gone as far as claiming that "any individual study must be considered only a single data point to be contributed to a future meta-analysis" (Schmidt, 1992); p. 1179). Without endorsing such an extreme viewpoint, we agree that a careful sifting and quantitative synthesis of the research conducted to date will help to advance the state of the field.

*1.2 Four general categories of meditation*

Before conducting a synthesis of the meditation neuroimaging literature, one must ask a central question: How are meditation practices to be categorized? There is growing consensus that at least three broad categories of meditation techniques can be discerned: *focused attention*, *open monitoring*, and *compassion* or *loving-kindness* practices (Cahn and Polich, 2006, Lutz et al., 2008b, Brewer et al., 2011, Vago and Silbersweig, 2012, Lippelt et al., 2014). Similar classification schemes appear to reach back centuries to early Eastern treatises on meditation (Harvey, 1990, Lodro, 1998, Goenka, 2000, Wangyal and Turner, 2011). Although a few broad categories do not begin to exhaust the subtleties of all contemplative practices (Niranjananada, 1993, Singh, 2002), there seems to be fair justification for a tentative division along these lines (Lutz et al., 2008b, Vago and Silbersweig, 2012). The main aim of this review and meta-analysis was to investigate whether some isomorphism (even if a complex one) can be discovered between psychological practices and brain recruitment (Cacioppo and Tassinary, 1990): that is, are these various categories of meditation, which differ markedly at the psychological level ($\Psi$), in fact characterized by distinctive patterns of brain activation and deactivation ($\Phi$)?

**1.2.1 Focused attention meditation.** *Focused attention* meditation involves directing attention to one specific object (e.g., the breath or a mantra) while monitoring and disengaging from extraneous thoughts or stimuli (Harvey, 1990, Hanh, 1991, Kabat-Zinn, 2005, Lutz et al., 2008b, Wangyal and Turner, 2011). Attention is directed to a particular object (most commonly the sensations associated with respiration), and whenever the mind wanders, attention is redirected to this target. With regular training, the ability to voluntarily control attention without being distracted appears to become progressively enhanced (e.g., (MacLean et al., 2010) and less effortful (Wallace, 1999, 2006, Brefczynski-Lewis et al., 2007). In particular, focused attention meditation may lead to three types of improvements: monitoring the locus of attention, disengaging from distraction, and shifting attention back to its intended target (Lutz et al., 2008b).

      **1.2.2 Mantra recitation meditation.** Focused attention meditation itself may be subdivided into distinct practices, or distinguished from practices that appear similar at first glance: in particular, *mantra recitation* meditation involves unique components (Travis, 2014). Mantra meditation – probably best known in the form of the widespread 'Transcendental Meditation' – involves the repetition of a sound, word, or sentence (spoken aloud or silently in one's head) with the goals of calming the mind, maintaining focus, and avoiding mind-wandering. While mantra meditation therefore clearly overlaps with other forms of focused attention in regard to its aims, it differs in that the object of focus is a voluntary verbal-motor production, rather than naturally arising body sensations (like the breath) or external physical objects (such as a point in space upon which the gaze is focused). Further, mantra meditation appears to be associated with neural correlates separate from other related forms of focused attention practice (e.g., (Lazar et al., 2000, Shimomura et al., 2008, Davanger et al., 2010, Tomasino et al., 2013) – although, in the absence of a meta-analysis, these differences remain only suggestive. Although mantra meditation could reasonably be placed within the general 'focused attention' category of practices, because of its unique elements (most notably its linguistic, verbal-motor component) we examined it separately here.

      **1.2.3 Open monitoring meditation.** *Open monitoring* practice typically begins with bringing attention to the present moment and impartially observing all mental contents

(thoughts, emotions, sensations, etc.) as they naturally arise and subside. A key element of this practice is possessing an open, accepting attitude toward, and learning to 'let go' of, mental content – neither resisting nor elaborating upon anything that surfaces in awareness (Harvey, 1990, Walker, 1995, Suzuki, 2003, Kabat-Zinn, 2005, Wangyal and Turner, 2011). In contrast to focused attention, then, in open monitoring meditation mental content is neither evaluated for relevance to a particular goal, nor is any content suppressed. This 'non-elaborative' mental stance cultivates a more present-centered awareness in tune with the moment-to-moment experience of the mind and body (Anālayo, 2003, Farb et al., 2007). Somatic and visceral body sensations are often a prominent feature of this present moment experience, and hence interoceptive and exteroceptive inputs generally receive greater processing in open monitoring than in focused attention practices (e.g., (Fox et al., 2012, Kerr et al., 2013). Open monitoring meditation can also sometimes serve as a platform from which practitioners can go on to enter subtler states of non-dual awareness (see section 1.3, below).

**1.2.4 Loving-kindness and compassion meditations.** *Loving-kindness* meditation (closely related to, but not identical with, *compassion* meditation) aims to deepen feelings of sympathetic joy for all living beings, as well as promote altruistic behaviors (Harvey, 1990, Gyatso and Jinpa, 1995, Kabat-Zinn, 2005, Lutz et al., 2008a). Typically, practitioners begin by generating feelings of kindness, love, and joy toward themselves, then progressively extend these feelings to imagined loved ones, acquaintances, strangers, enemies, and eventually all living beings (Harvey, 1990, Kabat-Zinn, 2005, Lutz et al., 2008a). *Compassion* meditation generally takes this practice a step further: practitioners imagine the physical and/or psychological suffering of others (ranging from loved ones to all humanity) and cultivate compassionate attitudes and responses to this suffering. Both practices share the long-term goals of enhancing sympathetic joy and empathy for pain, which are viewed as trainable skills that increase altruistic behavior (Gyatso and Jinpa, 1995, Lutz et al., 2008a). Although there are potentially important psychological differences between these practices, of the neuroimaging studies to date, the instructions have been comparable: participants have been asked to generate positive emotions (such as loving-kindness, well-wishing, or compassion) directed toward others and/or generated in a non-referential way (see the supplementary material for the exact instructions in each study). We therefore grouped the available studies together, reasoning that they were more similar than different: despite some difference in instructions and implementation of the practices, all studies involved a strong focus on cultivating positive affect directed outward. We acknowledge, however, that differences in the various stages of these practices (e.g., compassion for self, vs. a loved one, vs. a stranger, vs. non-referential compassion) might very well be resolved at the neural level by future research  Note, too, that the instructions involved in the studies investigated here do not exhaust the possibility of this type of practice, which can also involve complex visualizations and a focus on spiritual 'benefactors' (Makransky et al., 2012). Further research is clearly warranted to more thoroughly examine the many variants of loving-kindness and compassion meditation.

In one sense, loving-kindness and compassion meditations could also be considered a form of *focused attention* in that they often focus intensively on a single object (the person who is the target of the loving-kindness) and cultivate a consistent emotional tone (to the exclusion of other kinds of affect). Alternatively, when conducted in a non-referential, all-embracing way (extending one's compassion and joyful feelings to any being or object that arises in consciousness indiscriminately, or even cultivating such feelings without any object whatsoever), these practices could conceivably be considered a form of *open monitoring*. Nonetheless, given the strong emphasis on deliberate cultivation of joyful, altruistic, and empathetic emotions that is fairly unique to these practices, there seems sufficient justification for considering them as a separate class, worthy of investigation in its own right.

*1.3 Other forms of meditation*

It is important to note that these four putative categories of meditation are intended not as a final classification scheme, but merely as a first step toward delineating quantitative neurophysiological (Φ) correlates associated with relatively distinctive psychological (Ψ) practices (Cacioppo and Tassinary, 1990). There are several other major categories of meditation that have been investigated in some preliminary work, but that could not be examined meta-analytically due to a paucity of data. We nonetheless sought to examine and compare our meta-analytic results with these data by conducting a qualitative review of the results reported to date.

*Visualization* meditation, for instance, involves creating and sustaining complex mental imagery for extended periods of time. These visualizations typically involve mandalas or yantras (geometric designs) or deities or patron-teachers (Gyatso, 1981), and experts in these techniques have been shown to have enhanced visuospatial processing (Kozhevnikov et al., 2009). Nonetheless, only a single study of which we are aware has investigated the neural correlates of any form of visualization meditation (Lou et al., 1999), although loving-kindness and compassion practices can also sometimes involve important visualization components (Weng et al., 2013).

Another almost uninvestigated set of practices based in yogic traditions involve the *withdrawal of the senses* (*pratyahara* in Sanskrit; (Iyengar, 2005). These practices involve the deliberate detachment from or blocking of sensory inputs, including pain, with the eventual goal of transcending any narrow sense of self or personal identity. Although this type of practice shares many features of focused attention techniques (e.g., a highly selective attentional focus), and would traditionally be practiced alongside them (Iyengar, 2005), it could also be considered a practice in its own right. However, so far only one study has investigated neural correlates of this kind of technique (Kakigi et al., 2005).

*Non-dual awareness* practices are another example. In general, the aim of these practices is to dissolve or attenuate the boundary between subject and object (Josipovic, 2010, Travis and Shear, 2010b, Dunne, 2011). Although this kind of practice bears some resemblance to open monitoring meditation, it could be considered a distinct practice, and further investigation into its neural correlates would be a welcome follow-up to a recent, seminal study (Josipovic et al., 2011, Josipovic, 2014).

Finally, another suite of practices aims at manipulating or enhancing meta-awareness during various stages of sleep and dreaming. *Yoga nidra* (literally 'sleep yoga'), for instance, involves the deliberate dampening of sensory inputs to induce a 'hypnagogic' state (Mavromatis, 1987, Hori et al., 1994, Hayashi et al., 1999, Stenstrom et al., 2012), so as to enhance one's capacity to carry out visualization and relaxation practices (Saraswati, 1984). *Dream yoga* (*rmi-lam* in Tibetan; *svapnadarsana* in Sanskrit) practices aim at enhancing meta-awareness during dreaming (i.e., enhancing so-called 'lucid' dreaming;

(Gackenbach and LaBerge, 1988), and are particularly prevalent in the Tibetan Buddhist tradition (Norbu and Katz, 1992, Mullin, 1996, Rinpoche, 2004). Other techniques even aim at maintaining awareness during states of deep, dreamless sleep (Aurobindo, 2004, Sharma, 2012). We are aware of only a single study of these practices, however (Kjaer et al., 2002).

*1.4 Delineating reliable neural correlates of different meditation practices*

Only in about the past year have a sufficient number of studies been cumulatively reported to allow for a reliable meta-analysis of each major meditation type (including open monitoring and loving-kindness/compassion meditations). Even very recent meta-analyses have been forced by a paucity of empirical data to examine only focused attention and mantra recitation meditation (Tomasino et al., 2013), or to group all practice types together (Sperduti et al., 2012). Other meta-analyses have employed less theoretically or empirically grounded categorizations, such as basing meta-analytic contrasts on the religion of origin for a practice (Buddhism vs. Hinduism) as opposed to the phenomenological content, intended goals, and behavioral characteristics of the practice itself (Tomasino et al., 2014).

Here we present the first meta-analysis inclusive of four major categories of meditation, with a focus on elucidating whether ostensibly distinct forms of mental practice activate a common neural substrate (Sperduti et al., 2012) or instead show dissociable patterns of underlying neural activation and deactivation. Accordingly, we conducted four separate activation likelihood estimation (ALE) meta-analyses using 263 foci of peak activation from 26 independent PET and fMRI studies involving 32 unique experiments (Tables 1 and 2) to examine the patterns of brain activation associated with focused attention, mantra recitation, open monitoring, and loving-kindness/compassion meditation. Additionally, we examined the results for three categories of meditation that have only been investigated thus far in a single study each (Table 1). Because this paucity of data precluded a formal, quantitative meta-analysis, we qualitatively reviewed the results for each practice type and compared it to our other meta-analytic results (Table 1).

*Table 1.*   Seven meditation categories investigated via quantitative meta-analysis and qualitative review.

| Meditation Type | Contributing Experiments (*N*) | Contributing Foci A/D (Total) |
|---|---|---|
| Quantitative meta-analysis | | |
| Focused attention | 7 | 48/13 (61) |
| Mantra recitation | 8 | 63/19 (82) |
| Open monitoring | 10 | 45/29 (74) |
| Compassion/Loving-kindness | 6 | 36/4 (40) |
| **Totals** | 31 | 192/65 (257) |
| Qualitative review | | |
| Visualization | 1 | – |
| Sense-withdrawal | 1 | – |
| Non-dual awareness | 1 | – |

*Note.* Some studies examined multiple meditation types and therefore contributed to multiple analyses. This resulted in a total of 31 experiments (contrasts) from the 25 independent studies included in the quantitative meta-analyses for the first four categories (see Table 2 for details of included studies). Further, three other categories of meditation were also examined where results have been reported from only a single study to date. These investigations were not included in any formal, quantitative meta-analyses, but the results were nonetheless examined and compared with our meta-analytic findings in a qualitative fashion. A = activations; D = deactivations.

*1.5 Are the effects of meditation practices on brain function of any practical significance?*

Determining whether there are any consistencies in brain activation for various styles of meditation is only the first step. A logical next question is whether activations and deactivations associated with a given meditation technique have any practical significance. That is, what effect sizes characterize these differences, and do effect sizes vary across particular forms of meditation? Note, however, that practical significance is not a fully objective measure – typically a half-standard deviation difference is considered a notable effect, but this boundary is essentially arbitrary, and 'practical' significance does not necessarily translate to clinical significance or relevance (Rosnow and Rosenthal, 1989).

An additional benefit of calculating effect sizes is that they can be used to estimate the extent of publication bias in a field (Egger et al., 1997), a major and outstanding issue across the social and neurobiological sciences, including contemplative neuroscience. Because we found strong evidence for publication bias in morphometric (anatomical) neuroimaging studies of meditation in a recent review (Fox et al., 2014), we sought to address whether or not such a bias might be affecting *functional* neuroimaging studies of meditation as well. The second central aim of this meta-analytic synthesis was therefore to provide preliminary estimates of effect sizes in the functional neuroimaging of meditation in order to assess practical significance and the possibility of publication bias.

## 2. Meta-analytic methods

### *2.1 Literature review*

**2.1.1 Search strategy.** Three authors (KCRF, SN, and MG) searched MEDLINE (http://www.ncbi.nlm.nih.gov/pubmed/), Google Scholar (http://scholar.google.com), and PsycINFO (http://www.apa.org/pub/databases/psycinfo/index.aspx) for all papers containing the word 'meditation' since the first functional neuroimaging study of contemplative practices was published (Lou et al., 1999). These extensive lists of articles were then refined by searching within results for studies that contained any of the words or phrases 'magnetic resonance imaging', 'MRI,' 'neuroimaging,' or 'brain' within the title or abstract. Of the remaining results, every abstract was consulted to see whether the study actually employed functional neuroimaging methods to study meditation. The reference lists of each study found, as well as those of several major reviews (e.g., (Holzel et al., 2011, Vago and Silbersweig, 2012)), were also consulted to ensure that no studies were missed.

**2.1.2 Study inclusion and exclusion criteria.** All studies using functional neuroimaging to investigate some form of meditation were considered. By 'functional neuroimaging' we mean functional magnetic resonance imaging (fMRI) or positron emission tomography (PET) studies that can provide details about the specific location of brain activations and deactivations. Using the search strategy detailed above, a total of 78 studies served as the initial pool of data to be reviewed (see Table 2 for included studies, and Table S1 for excluded studies).

In order to ensure the most rigorous results possible, stringent inclusion criteria were applied. Studies were required to: (i) report specific peak foci of activation in either

Talairach or Montreal Neurological Institute (MNI) space; (ii) include a reasonable sample size (i.e., case studies of single subjects were excluded); (iii) involve participants who actually practiced meditation, be they novices or long-term practitioners (as opposed to, for example, studies involving only self-report scales of 'dispositional' mindfulness); (iv) involve actual meditation *during the scanning session* (as opposed to having meditation practitioners engage in other tasks without some explicit contemplative element); and (v) involve healthy, non-clinical populations. Finally, (vi) only reports published in peer-reviewed scientific journals were included (results from conference abstracts, presented talks, dissertations, etc., were excluded). For further details on excluded studies, see the Supplementary Material Methods.

Ultimately, about one third (25 of 78) of the examined studies were included in our meta-analyses (Table 2). Given that some of these studies examined multiple categories of meditation, we investigated a total of 31 separate 'experiments' or contrasts of interest spread across the four major meditation categories.

**2.1.3 Study classification.** Two coders with longtime personal meditation experience (authors KCRF and MLD) independently classified all included results as involving *focused attention*, *open monitoring*, *compassion* or *loving-kindness*, or *mantra recitation* meditation. Concordance was nearly perfect: a single disparity was resolved after further discussion. Several studies employed multiple practice types across different sessions; in these cases, results from each session were coded separately and considered individual 'experiments.' Therefore, the number of 'experiments' exceeded the number of independent studies actually included in the meta-analysis (see Tables 1 and 2). All classifications for included studies are summarized in Table 2.

Further, one study each was found that examined *visualization* (Lou et al., 1999), *sense-withdrawal* (Kakigi et al., 2005), and *non-dual awareness* (Josipovic et al., 2011) forms of meditation. Again, both coders (KCRF and MLD) independently classified and agreed on these categorizations. Due to there being only a single study for each category, these reports were not included in any formal meta-analyses. Nevertheless, their results were qualitatively reviewed and considered in comparison to our meta-analytic results for the other four general meditation styles (see Table 1).

In virtually all cases, classification was straightforward and mirrored the classifications described by the authors of the studies in their methods. For instance, phrases such as 'concentration meditation' were categorized under the focused attention category; mantra meditation studies were easily categorized based on descriptions of mantra recitation in the methods; and so on. Because of the prevalence of classification schemes involving the four categories described above (e.g., (Cahn and Polich, 2006, Lutz et al., 2008b, Travis and Shear, 2010a, Vago and Silbersweig, 2012), many studies in fact explicitly described their methods using these same (or highly similar) terms.

*Table 2.*   Studies included in the meta-analyses ($n$ = 25).

| Study | Practice | N (M/C) | Tradition | Expertise | Average Experience |
|---|---|---|---|---|---|
| Lou et al. (1999) | OM + LK | 9/– | Yoga Nidra | LTP | >5 yrs |
| Lazar et al. (2000) | MR | 5/– | Kundalini | LTP | 4 yrs |
| Brefczynski-Lewis et al. (2007) | FA | 14/16 | Tibetan Buddhist | LTP | 19000 hrs & 44000 hrs |
| Farb et al. (2007) | OM | 20/– | MBSR | STT | ~42 hours |
| Lutz et al. (2008) | LK | 16/16 | Tibetan Buddhist | LTP | 45 ± 12.7 yrs |
| Shimomura et al. (2008) | MR | 8/– | Pure Land Buddhism | LTP | >10 years |
| Lutz et al. (2009) | LK | 10/12 | Tibetan Buddhist | LTP | 40 ± 9.6 yrs |
| Davanger et al. (2010) | MR | 4/– | Acem Meditation | LTP | >23 yrs |
| Engström et al. (2010) | MR | 8/– | Acem Meditation and Kundalini Yoga | LTP | 1.17 yrs |
| Manna et al. (2010) | FA + OM | 8/– | Vipassana (Theravada - Thai Forest Tradition) | LTP | 15750 hrs |
| Brewer et al. (2011) | FA + OM + LK | 12/13 | Vipassana (Mindfulness) | LTP | 10565 ± 5148 hrs |
| Gard et al. (2011) | OM | 17/17 | Vipassana | LTP | 5979 ± 5114 |
| Ives-Deliperi et al. (2011) | OM | 10/– | MBSR | LTP | >4 yrs |
| Kalyani et al. (2011) | MR | 12/– | *Om* mantra recitation | LTP + STT | Not reported |
| Taylor et al. (2011) | OM | 12/10 | Zen | LTP + STT | LTP: >1,000 hrs STT: 1 wk |
| Wang et al. (2011) | MR | 10/– | Kundalini yoga | LTP | ~20,000 hrs |
| Dickenson et al. (2012) | FA | 31/– | Mindfulness/MBSR | STT | No previous experience; brief meditation induction only |
| Hasenkamp et al. (2012) | FA | 14/– | Various | LTP | 1386 ± 1368 hrs |
| Lee et al. (2012) | FA + LK | 22/22 | Theravada | LTP | FA: 5249 ± 6192 hrs LK: 7492 ± 6681 hrs |
| Farb et al. (2013) | OM | 20/16 | MBSR | STT | 49.2 ± 4.3 hrs |
| Guleria et al. (2013) | MR | 14/– | Soham Meditation | LTP | 2088 ± 320 hrs |
| Lutz et al. (2013) | OM | 14/14 | Tibetan Buddhist | LTP | 27000 ± 12500 hrs |
| Weng et al. (2013) | LK | 20/21 | Tibetan Buddhist | STT | 5.9 ± 0.5 hrs |
| Lutz et al. (2014) | OM | 24/22 | Mindfulness/MBSR | STT | Variable |
| Xu et al. (2014) | FA + MR | 14/– | Acem Meditation | LTP | 27 ± 9 yrs |

All studies employed fMRI except for Lou et al. (1999), which employed PET. C: controls; FA: focused attention; LK: loving-kindness or compassion meditation; LTP: long-term practitioners; M: meditators; MBSR: mindfulness-based stress reduction; MR: mantra recitation; *N*: sample size; OM: open monitoring; STT: short-term training.

*2.2 The question of baselines, control conditions, and tasks engaged in during meditation*

One potential problem is that even studies examining a similar form of meditation might compare a meditation practice to very different baseline or comparison tasks and conditions. However, collapsing across numerous different baselines or control conditions is a common (in fact, usually inevitable) practice in meta-analyses of functional neuroimaging studies (Turkeltaub et al., 2002, Wager et al., 2003, Wager et al., 2007, Wager et al., 2009, Caspers et al., 2010, Yarkoni et al., 2010, Tomasino et al., 2013)**.** It is by no means a problem limited to the meditation literature. Aside from the heterogeneity of conditions across studies, meditation is also often investigated *during* a given task or form of stimulus presentation. Although typically these tasks or stimuli are equally present in the baseline or control condition (and therefore, in principle, should not unduly influence the meditation state results), the potential for our results representing interactions between meditation and the given task employed should be kept in mind. These issues are examined further in the Discussion.

The goal of the present synthesis was to determine, *irrespective* of any incidental differences in comparison or baseline conditions and tasks: (i) whether differing meditation practices tend to reliably recruit differentiable neural networks; (ii) if the differences in brain activation show practically significant effect sizes; and lastly, (iii) whether distinct practices exhibit differing mean effect sizes (i.e., is a given style of meditation associated with greater differences in brain activity than others?).

*2.3 Reporting and classification of results*

All peak voxel coordinates are reported in Montreal Neurological Institute (MNI) space. For consistency, for all studies where peak voxels were originally reported in Talairach coordinates, we used the WFU Pickatlas software package (Maldjian et al., 2003) to perform a nonlinear transformation of Talairach coordinates to MNI space. For meta-analytic ALE results, region classifications follow those indicated in the Multi-Image Analysis GUI ('Mango') image-viewing software (UT Health Science Center Research Imaging Institute) used to display our findings (see Results below). For additional precision, the Duvernoy neuroanatomical atlas was also consulted (Duvernoy et al., 1991).

*2.4 Activation likelihood estimation (ALE) meta-analysis*

**2.4.1 General methods.** We used a quantitative, random-effects meta-analytic method known as activation likelihood estimation (ALE; (Turkeltaub et al., 2002, Laird et al., 2005, Eickhoff et al., 2009, Eickhoff et al., 2012, Turkeltaub et al., 2012) implemented in the software program GingerALE 2.1.1 (San Antonio, TX: UT Health Science Center Research Imaging Institute). The most recent ALE algorithm tests for above-chance clustering of peak foci from different experiments included in the meta-analysis (Eickhoff et al., 2009, Eickhoff et al., 2012) by comparing actual activation foci locations/clustering with a null distribution that includes the same number of peak foci distributed randomly throughout the brain. Included activation foci were smoothed using a full-width half maximum (FWHM) Gaussian kernel dependent on the total sample size of the experiment from which foci were drawn (larger sample -> smaller smoothing kernel – empirically determined by (Eickhoff et al., 2009, Eickhoff et al., 2012). Resulting statistical maps show clusters where convergence between foci of activation or deactivation is greater than would be expected by chance (i.e., if foci from each experiment were distributed independently).

**2.4.2 Primary neuroimaging meta-analyses.** We meta-analyzed a total of 257 peak foci of activation or deactivation drawn from 25 studies (Table 2), involving 31separate 'experiments' or contrasts (summarized in Table 1). In order to retain a maximal amount of information, this primary meta-analysis collapsed data from between-group (long-term practitioners vs. novices or controls) and repeated-measures (the same practitioners pre- and post-meditation training) designs. Although only a small amount of data came from short-term training (STT) investigations (6 of 25 studies included in the meta-analysis), nonetheless this difference presents a potentially confounding factor. We therefore addressed this issue by performing supplemental meta-analyses that included only investigations of long-term practitioners (see Supplementary Materials).

Several coordinate foci ($n$ = 7) from the included studies fell outside of the brain mask templates used with the GingerALE meta-analysis software (1 focus for focused attention activations; 4 foci for open monitoring activations; and 2 foci for loving-kindness activations). This is a normal occurrence when using exclusive meta-analytic template masks. The total number of foci included in the final meta-analyses was therefore 250.

Statistical maps were thresholded using a false discovery rate (FDR; (Genovese et al., 2002) of $q$ = .05 and a cluster threshold of $k$ = 100 mm$^3$. The meta-analytic software offers suggested minimum cluster thresholds depending on the number of studies and foci entered. These recommendations centered around 100 (generally, between 88 and 132 mm$^3$), but differed slightly in each of our eight meta-analyses (activations and deactivations for each of four meditation categories). For consistency, we therefore set a standard value of 100 mm$^3$ for all analyses. Note that the cluster threshold is essentially arbitrary and does not affect the underlying meta-analytic analyses or results; it merely determines which meta-analytic clusters are deemed 'significant' and therefore to be reported in tables of results. It does not affect the underlying meta-analytic results, which display all meta-analytic activations and deactivations regardless of cluster threshold. The full meta-analytic data are visible in our supplementary results figures.

To display our results, we used the 'Colin' template brain images suggested for use with GingerALE 2.1.1, displayed in the 'Mango' software package (San Antonio, TX: UT Health Science Center Research Imaging Institute). Maps were computed separately for activations and deactivations for each meditation category. Activation and deactivation maps were then concatenated onto a single template brain image for visualization purposes. The creation of the final figures for presentation here was performed in the MRIcron software package (http://www.mccauslandcenter.sc.edu/mricro/mricron/index.html) (Rorden et al., 2007).

*2.5 Effect size meta-analysis*

In addition to determining which brain regions were consistently activated by various meditation practices, we sought to evaluate the magnitude of these differences (i.e., their effect sizes; (Cohen, 1992, Lipsey and Wilson, 2001, Sedlmeier et al., 2012). In all, 17 of the 25 studies that met our inclusion criteria also provided necessary and sufficient data to allow reliable calculation of effect sizes.

Although the calculation and comparison of effect sizes in neuroimaging studies faces a number of challenges, simple null-hypothesis significance tests and *p*-values have come under harsh criticism as well (Schmidt, 1992, Schmidt and Hunter, 1997, Cumming, 2012, 2013, Hupé, 2015). The meaning and relevance of null-hypothesis significance testing and associated statistics (such as *t* and *F* statistics) has become so hotly contested that some journals have outright banned reporting of such values (Savalei and Dunn, 2015, Trafimow and Marks, 2015).

Bearing in mind these considerations, we concluded that effect sizes are worth calculating and reporting in meta-analyses of neuroimaging data, so long as these serious limitations are noted and attempts are made to correct for them. The many relevant issues are discussed at length in the Supplementary Methods section.

Briefly, our procedure for applying these corrections was as follows: After calculating effect sizes directly from the peak or *maximum t* or *F* statistics reported in the original studies, we then adjusted these values downward to approximate more conservative estimates of *t* or *F* statistics for the entire *cluster* of significant difference. Next, we further deflated these values to account for the inflationary bias of effect sizes derived only from results that have exceeded stringent statistical thresholds (Schmidt, 1992). In these ways, we aimed to provide effect size estimates that more closely approximate true effect sizes. It should be noted, however, that we neither claim nor expect that this procedure is ideal. It is clear that some deflation of reported values is necessary (Schmidt, 1992, Yarkoni, 2009, Yarkoni et al., 2010, Hupé, 2015), but the best method and optimal degree of deflation remains unclear. As such, our methods are a preliminary effort to produce effect sizes that are more accurate than those produced by calculations solely based on *t*-statistics reported in the literature, but should not by any means be considered a definitive approach. For full details, as well as specific formulas used, see the Supplementary Methods.

*2.6 Estimating publication bias in meta-analytic results*

The bias toward publication of only positive (i.e., non-null) results is a serious concern (the 'file drawer' problem; c.f. (Rosenthal, 1979). We constructed a funnel plot (scatterplot of effect size against sample size) to test for potential publication bias in our sample of studies (Egger et al., 1997). Effect sizes were calculated as described above and in the Supplementary Methods, and plotted against total sample size (meditators + controls). For detailed discussion of funnel plots see Egger and colleagues (1997). For other examples of their use in meta-analyses of meditation and further discussion see Fox and colleagues (2014) and Sedlmeier and colleagues (2012).

## 3. Results I: Neuroimaging meta-analysis

*3.1 Focused attention meditation*

Meta-analysis of focused attention studies resulted in 2 significant clusters of activation, both in prefrontal cortex (Table 3; Fig. 2). Activations were observed in regions associated with the voluntary regulation of thought and action, including the premotor cortex (BA 6; Fig. 2e) and dorsal anterior cingulate cortex (BA 24; Fig. 2a). Slightly sub-threshold clusters were also observed in the dorsolateral prefrontal cortex (BA 8/9; Fig. 2c) and left mid insula (BA 13; Fig. 2e); we display these somewhat sub-threshold results here because of the obvious interest of these findings in practices that involve top-down focusing of attention, typically focused on respiration. We also observed clusters of deactivation in regions associated with episodic memory and conceptual processing, including the ventral posterior cingulate cortex (BA 31; Fig. 2d) and left inferior parietal lobule (BA 39; Fig. 2f). A detailed series of slices covering the entire brain is presented in Fig. S1.

A supplementary meta-analysis, excluding studies that investigated only practitioners with short-term training, yielded nearly identical results (see Table S2).

*Table 3.* Activations and deactivations associated with focused attention meditation

| Region | Cluster Size (mm³) | Side | Peak Coordinates (x, y, z) | Peak ALE value |
|---|---|---|---|---|
| **Activations** | | | | |
| Premotor cortex | 712 | L | -36, 6, 56 (BA 6) | 0.0144 |
| Dorsal anterior/mid cingulate cortex | 280 | M | 2, 12, 32 (BA 24) | 0.0112 |
| **Deactivations** | | | | |
| Posterior cingulate cortex | 152 | M | -6, -60, 18 (BA 30) | 0.0071 |
| Inferior parietal lobule | 144 | L | -48, -72, 30 (BA 39) | 0.0071 |

*Figure 2.* Peak activations and deactivations associated with focused attention meditation

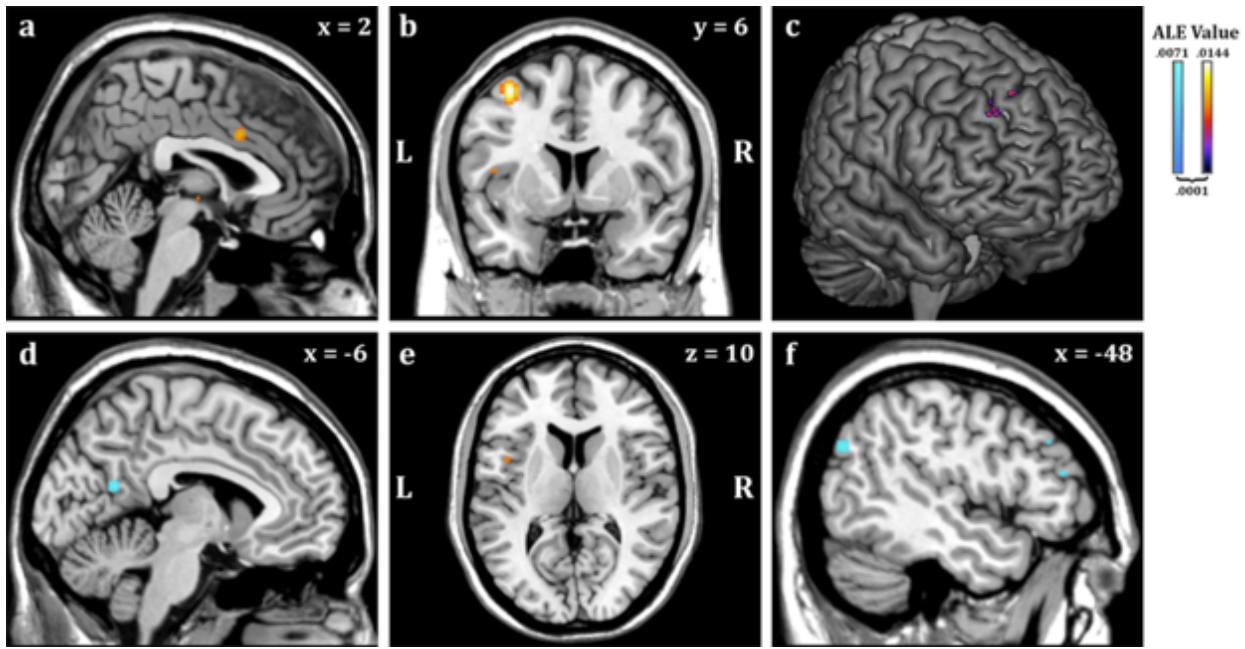

(a) Dorsal anterior/mid cingulate cortex (b) Premotor cortex/posterior dorsolateral prefrontal cortex. (c) Sub-threshold cluster in dorsolateral prefrontal cortex. (d) Posterior cingulate cortex. (e) Sub-threshold cluster in mid-insula. (f) Inferior parietal lobule. Warm colors: activations; cool colors: deactivations. See the supplementary materials for a detailed series of slices covering the entire brain (Fig. S1).

*3. 2 Mantra recitation meditation*

A meta-analysis of mantra recitation studies revealed seven significant activation clusters (Table 4). Activations were observed in regions associated with planning and executing voluntary motor output, including the posterior dorsolateral prefrontal cortex/left premotor cortex (BA 6/8; Fig. 3d), pre-supplementary motor cortex, supplementary motor cortex (BA 6; Fig. 3a), and putamen/lateral globus pallidus (Fig. 3b). Other activations were observed in regions associated with visual processing and mental imagery including the fusiform gyrus (Fig. 3c), cuneus (BA 18), and precuneus (BA 7) (Fig. S2). There was also a small (but non-significant) cluster located within Broca's area (BA 44/6), consistent with the verbal component of mantra recitation. A single cluster of deactivation was observed (Table 4) in the left anterior insula (BA 13)/claustrum (Fig. 3b), a region associated with processing viscero-somatic body signals. A detailed series of slices covering the entire brain is presented in Fig. S2.

A supplementary meta-analysis, excluding studies that investigated only practitioners with short-term training, yielded nearly identical results. The only notable difference was the disappearance of the single deactivation cluster in the anterior insula (see Table S3). Although ALE implements controls for sample size, another concern was the inclusion of two studies with very small sample sizes – Lazar et al. (2000) with $n$ = 5, and Davanger et al. (2010) with $n$ = 4 – which might still have unduly influenced the results. We therefore also executed our analysis with data from these two studies excluded. This had no effect on the pattern of deactivations. All remaining clusters of activation were the same, but two clusters disappeared from the results: the putamen/lateral globus pallidus, and the small (112 mm$^3$) cluster in the supplementary motor area.

*Table 4.* Activations and deactivations associated with mantra recitation meditation

| Region | Cluster Size (mm³) | Side | Peak Coordinates (x, y, z) | Peak ALE value |
|---|---|---|---|---|
| **Activations** | | | | |
| Premotor cortex | 896 | L | -30,4, 60 (BA 6) | 0.0137 |
| Supplementary motor area | 584 | M | 0, 12, 54 (BA 6) | 0.0111 |
| | 112 | M | -4, 2, 68 (BA 6) | 0.0080 |
| Putamen/Lateral globus pallidus | 368 | R | 28, -16, -6 | 0.0111 |
| Fusiform gyrus | 248 | R | 40, -26, -30 (BA 20/36) | 0.0093 |
| Cuneus | 160 | R | 24, -86, 26 (BA 18) | 0.0082 |
| Precuneus | 152 | L | -14, -56, 54 (BA 7) | 0.0082 |
| **Deactivations** | | | | |
| Anterior insula | 192 | L | -28, 25, -7 (BA 13) | 0.0075 |

*Figure 3.* Peak activations and deactivations associated with mantra recitation meditation

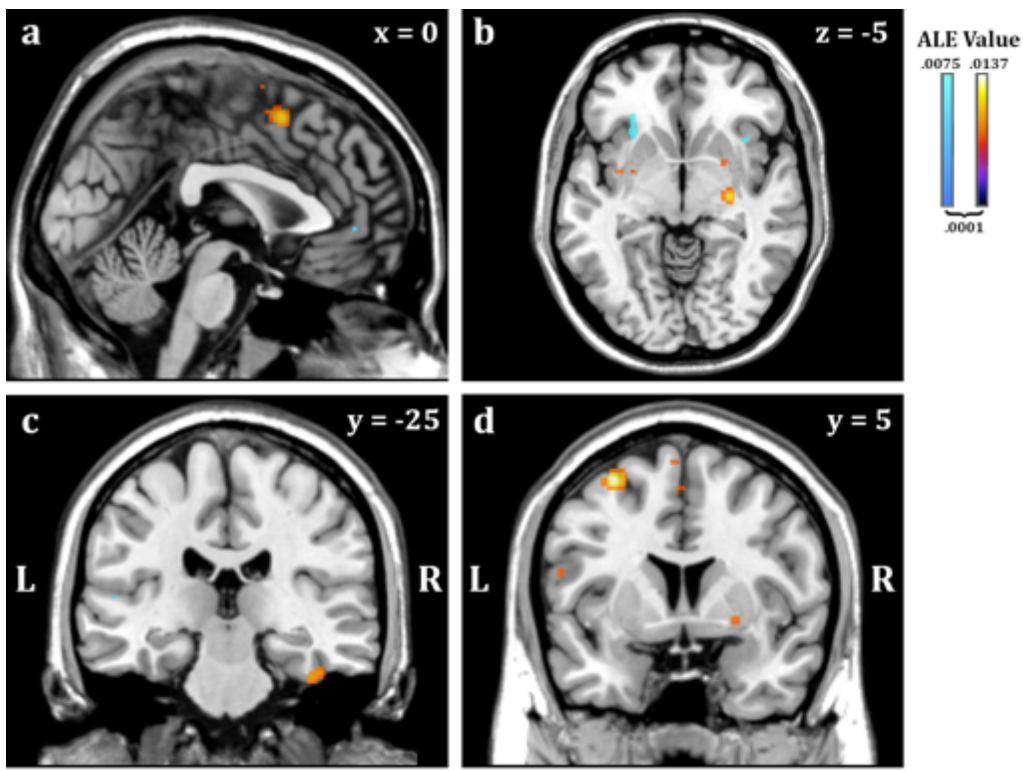

(a) Supplementary motor area. (b) Bilateral deactivations in the anterior insula (blue) and activations in the right basal ganglia (putamen and globus pallidus). (c) Fusiform gyrus. (d) Premotor cortex. Warm colors: activations; cool colors: deactivations. See the supplementary materials for a detailed series of slices covering the entire brain (Fig. S2).

*3.3 Open monitoring meditation*

Meta-analysis of open monitoring studies showed five significant clusters of activation (Table 5). One cluster was located in the insula (BA 13; Fig. 4d), consistent with awareness of ongoing viscero-somatic body signals. Other significant clusters were observed in regions associated with the voluntary control of action, including the left inferior frontal gyrus (BA 44/45; Fig. 4b), pre-supplementary motor area (BA 32/6; Fig. 4a), supplementary motor area (BA 6; Fig. 4a,c), and premotor cortex (BA 6; Fig. 4 c). Smaller (non-significant) clusters were observed in the rostrolateral prefrontal cortex (BA 10) and mid-dorsolateral prefrontal cortex (BA 9/46) — regions associated with cognitive control and metacognitive awareness. We observed a single significant cluster of deactivation (Table 5) in the right thalamus (Fig. 4d). A detailed series of slices covering the entire brain is presented in Fig. S3.

A supplementary meta-analysis, excluding studies that investigated only practitioners with short-term training, yielded similar results (Table S4). Some of the more salient differences are outlined in the Discussion, below (also, compare Table 5 with Table S4).

*Table 5.* Activations and deactivations associated with open monitoring meditation

| Region | Cluster Size (mm³) | Side | Peak Coordinates (x, y, z) | Peak ALE value |
|---|---|---|---|---|
| **Activations** | | | | |
| Supplementary motor area | 824 | M | -6, 4, 60 (BA 6) | 0.0165 |
| Dorsal anterior cingulate cortex/ pre-supplementary motor area | 368 | M | -6, 18, 44 (BA 32/6) | 0.0110 |
| Insular cortex (mid/anterior) | 360 | L | -44, 10, 4 (BA 13) | 0.0111 |
| Inferior frontal gyrus | 200 | L | -50, 16, 14 (BA 44/45) | 0.0092 |
| Premotor cortex | 104 | L | -44, 10, 46 (BA 6) | 0.0089 |
| **Deactivations** | | | | |
| Thalamus (pulvinar) | 112 | R | 17, -24, 11 | 0.0084 |

*Figure 4.* Peak activations and deactivations associated with open monitoring meditation

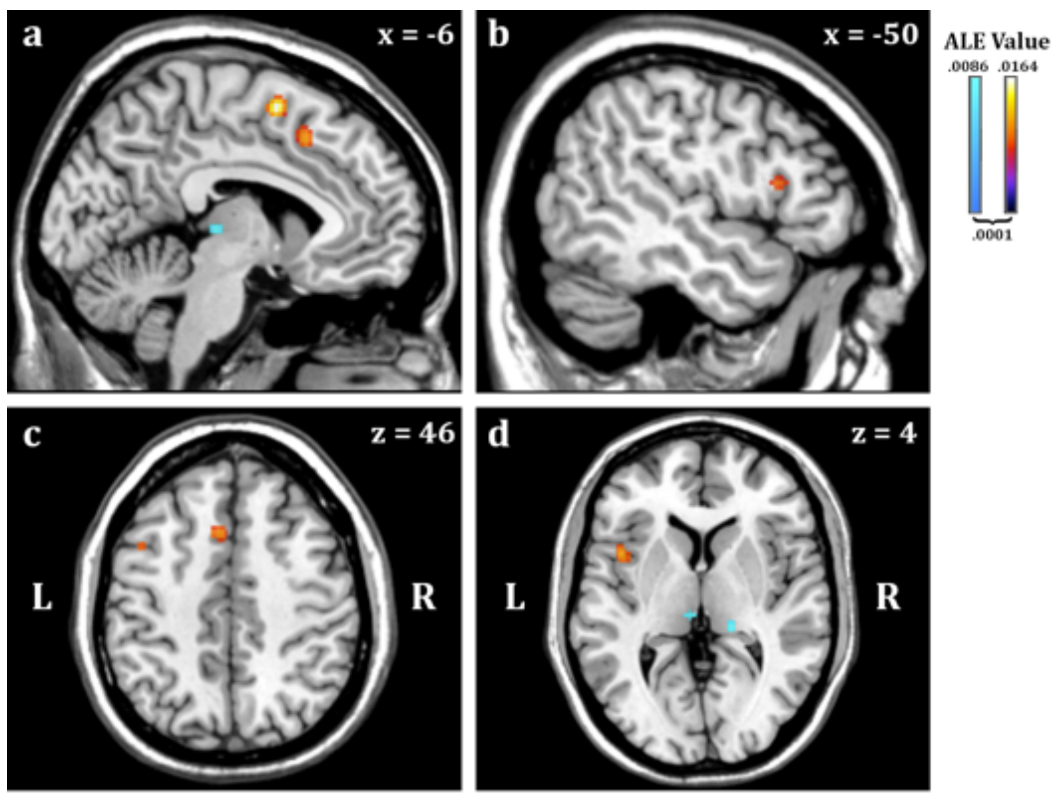

(a) Supplementary motor area (larger, more superior cluster) and dorsal anterior cingulate cortex/pre-supplementary motor area. (b) Inferior frontal gyrus. (c) Premotor cortex (smaller cluster to left), and supplementary motor area also visible again (center). (d) Activation in mid/anterior insula and bilateral deactivations in the thalamus. Warm colors: activations; cool colors: deactivations. L: left; R: right. See the supplementary materials for a detailed series of slices covering the entire brain (Fig. S3).

*3.4 Loving-kindness and compassion meditation*

Meta-analysis of loving-kindness and compassion meditation revealed significant activation clusters in 3 regions (Table 6). Activations were observed in regions associated with awareness of bodily sensations and feelings including the right anterior insula/frontal operculum (BA 13; Fig. 5a) and secondary somatosensory areas extending into the anterior inferior parietal lobule (BA 2/40; Fig. 5b). Additionally, activation was observed near the parieto-occipital sulcus (BA 23/31; Fig. 5c). No significant deactivations were observed. A detailed series of slices covering the entire brain is presented in Fig. S4.

A supplementary meta-analysis, excluding studies that investigated only practitioners with short-term training, yielded nearly identical results. The only significant difference was the appearance of a small cluster of activation in the left somatosensory cortices (see Table S5).

*Table 6.* Activations and deactivations associated with loving-kindness and compassion meditation

| Region | Cluster Size (mm³) | Side | Peak Coordinates (x, y, z) | Peak ALE value |
|---|---|---|---|---|
| **Activations** | | | | |
| Anterior insula | 512 | R | 38, 22, 14 (BA 13) | 0.0135 |
| Parieto-occipital sulcus | 344 | R | 24, -60, 18 (BA 23/31) | 0.0115 |
| Somatosensory cortices/Inferior parietal lobule | 320 | R | 54, -26, 30 (BA 2/40) | 0.0101 |

*Figure 5.*   Peak activations associated with loving-kindness and compassion meditation

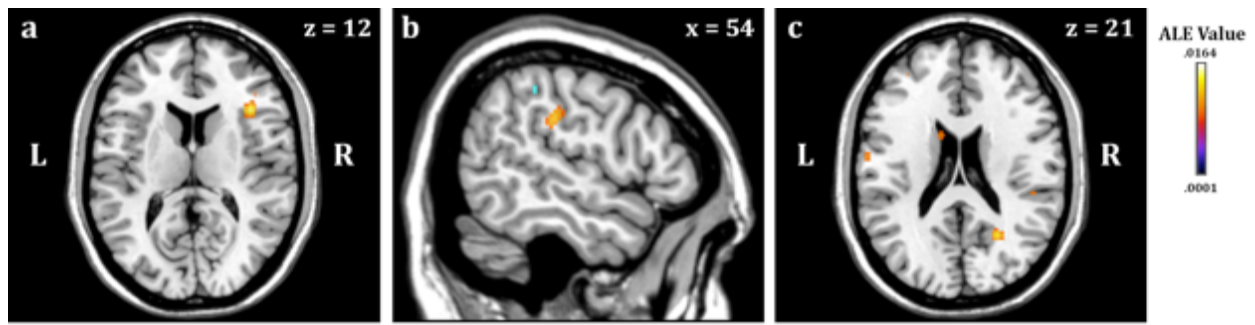

(a) Right anterior insula/frontal operculum. (b) Right somatosensory cortices. (c) Parieto-occipital sulcus. L: left; R: right. See the supplementary materials for a detailed series of slices covering the entire brain (Fig. S4).

## 4. Results II: Effect size meta-analysis

As detailed in the Supplementary Methods, we derived effect sizes (Cohen's *d*) from studies that reported *t* or *F* statistics, and then adjusted these raw effect sizes for the severe inflationary bias that influences effect sizes in all neuroimaging research. From our adjusted effect sizes for each individual reported result, we then calculated mean positive (+) and negative (−) effect sizes for each study (and for each different practice within a study, if more than one was examined). These results are reported in Table 7. We then pooled these data to calculate mean (+) and (−) effect sizes for each type of practice (Table 8). Recall that in total, two deflation coefficients were employed to adjust for suspected inflation bias. Coefficients of 0.697 and 0.60, for a net coefficient of 0.4182, were applied to the effects calculated from the primary reports. Therefore, in order to obtain the unadjusted values, simply multiply our effect size results by 2.391. For details see Methods section 2.5.

*Table 7.* Summary of adjusted mean effect sizes for each study and practice type.

| Study | Sample (*N*) | Practice | Mean (+) effect size (Cohen's *d*) | Mean (−) effect size (Cohen's *d*) |
|---|---|---|---|---|
| Brefczynski-Lewis et al. (2007) | 30 | FA | 0.51 | – |
| Lutz et al. (2008) | 32 | LK | 0.33 | – |
| Lutz et al. (2009) | 22 | LK | 0.56 | – |
| Davanger et al. (2010) | 4 | MR | 1.48 | – |
| Manna et al. (2010) | 8 | FA | 1.56 | -1.29 |
| | | OM | 1.25 | – |
| Brewer et al. (2011) | 25 | FA | – | -0.71 |
| | | OM | – | -0.58 |
| | | LK | – | -0.68 |
| Gard et al. (2011) | 34 | OM | 0.46 | – |
| Ives-Deliperi et al. (2011) | 10 | OM | 2.03 | -2.01 |
| Kalyani et al. (2011) | 12 | MR | – | -1.11 |
| Taylor et al. (2011) | 22 | OM | – | -0.79 |
| Dickenson et al. (2012) | 31 | FA | 0.49 | – |
| Hasenkamp et al. (2012) | 14 | FA | 0.67 | – |
| Lee et al. (2012) | 44 | FA | 0.56 | – |
| | | LK | 0.54 | – |
| Farb et al. (2013) | 36 | OM | 0.71 | – |
| Lutz et al. (2013) | 28 | OM | 0.58 | -0.62 |
| Weng et al. (2013) | 41 | LK | 0.36 | – |
| Lutz et al. (2014) | 46 | OM | 0.51 | -0.45 |

*Table 8.* Summary of adjusted mean effect sizes by meditation practice type

| Meditation Type | Contributing Experiments (*N*) | Mean (+) effect size (Cohen's *d*)[a] | Mean (−) effect size (Cohen's *d*)[a] |
|---|---|---|---|
| Focused attention | 6 | 0.60 ± 0.05 | -0.85 ± 0.09 |
| Mantra recitation | 2 | 1.48 | -1.11 |
| Open monitoring | 8 | 0.68 ± 0.12 | -0.69 ± 0.14 |
| Loving-kindness/compassion | 5 | 0.44 ± 0.01 | -0.68 |
| **All practices combined** | **21** | ***M* = 0.59 ± 0.02** | ***M* = -0.74 ± 0.09** |

Mean effect sizes ± 95% confidence intervals. Note that the 'mean' effect sizes for mantra recitation are based on only a single study each, as are negative effect sizes for loving-kindness/compassion meditations (cf. Table 7). Accordingly, no confidence intervals are provided since there was no variance in the estimates. Note that a total of 21 'experiments' were analyzed, from 17 independent studies of the 25 studies originally included in the neuroimaging meta-analysis. The remaining studies did not provide statistical data that allowed for calculation of Cohen's *d*.

*Figure 6.* Funnel plot showing weighted means of both negative and positive effect sizes

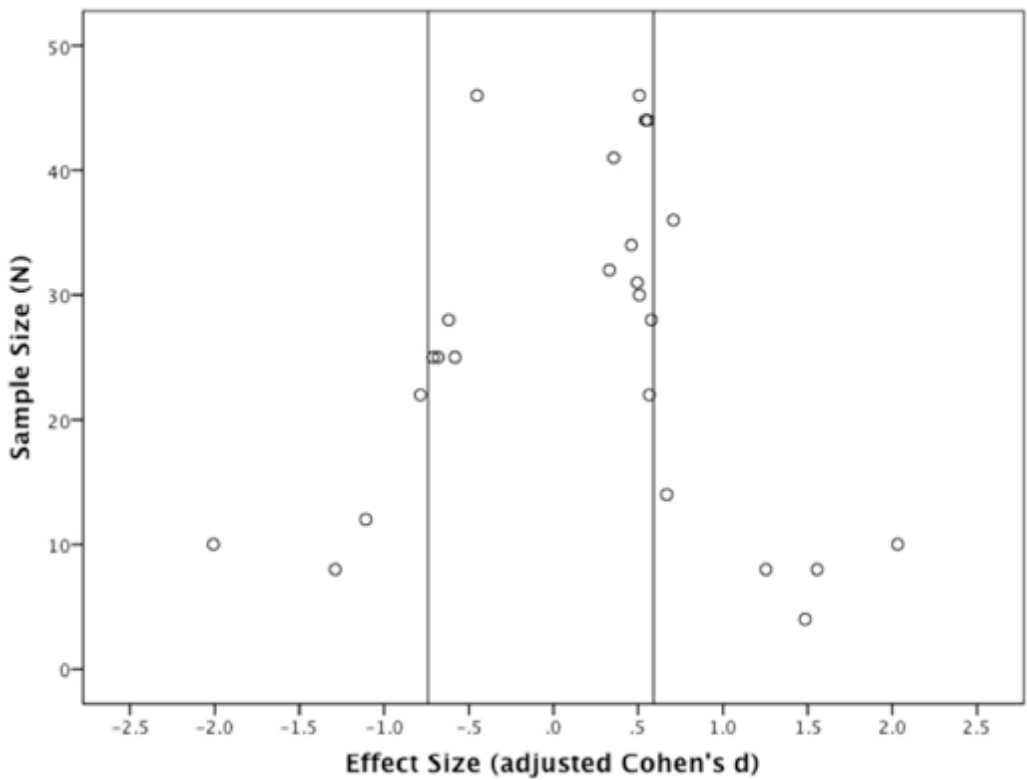

The two anchor lines represent the means for all (−) and (+) effect sizes (-0.74 and +0.59, respectively) weighted by sample size. Positive effect sizes relate to brain activations during meditation, whereas negative effect sizes represent brain deactivations. See Tables 7 and 8 for details.

## 5. Discussion, future directions, and conclusions

*5.1 Overview*

The ensuing discussion is divided into three broad sections:

(i) First, we outline and interpret the various results from our meta-analysis and qualitative review of neuroimaging findings (sections 5.2-5.8). We synthesize and simplify all of these results in a single summary figure (Fig. 7). We focus on the clusters that attained statistical significance in our meta-analysis, but it should be recalled that significance and cluster thresholds are ultimately arbitrary and should not restrict discussion. Some 'significant' clusters are difficult to understand or explain, whereas some *non*-significant clusters dovetail extremely well with the stated techniques and goals of a given form of meditation. We therefore endeavor to offer as integrated and impartial a discussion as possible, including consideration of meta-analytic activations that may not have exceeded our cluster thresholds. We also integrate the limited findings from the three meditation categories that could not be included in the quantitative meta-analysis, for the same reasons: preliminary data, even if tentative, is still informative and relevant to our discussion. In a similar vein, the figures in our Results section, for the sake of clarity, focus on the most reliable, consistent clusters for each category of meditation. However, we have presented detailed series of slices in our supplementary materials in order to clearly visualize *all* activations associated with a given meditation type (not just those that exceeded our cluster and significance thresholds; see Figs. S1-S4).

(ii) Second, we discuss the results of our effect size meta-analysis (section 5.9), and its implications for the 'practical significance' of the functional neural effects associated with meditation. We also discuss the possibility of publication bias.

(iii) Finally, several sections (5.10-5.13) outline the limitations of our meta-analysis, the relevance of these results to clinical conditions, and the implications for future avenues of research.

*Figure 7*. Simplified summary of all meta-analytic activations and deactivations.

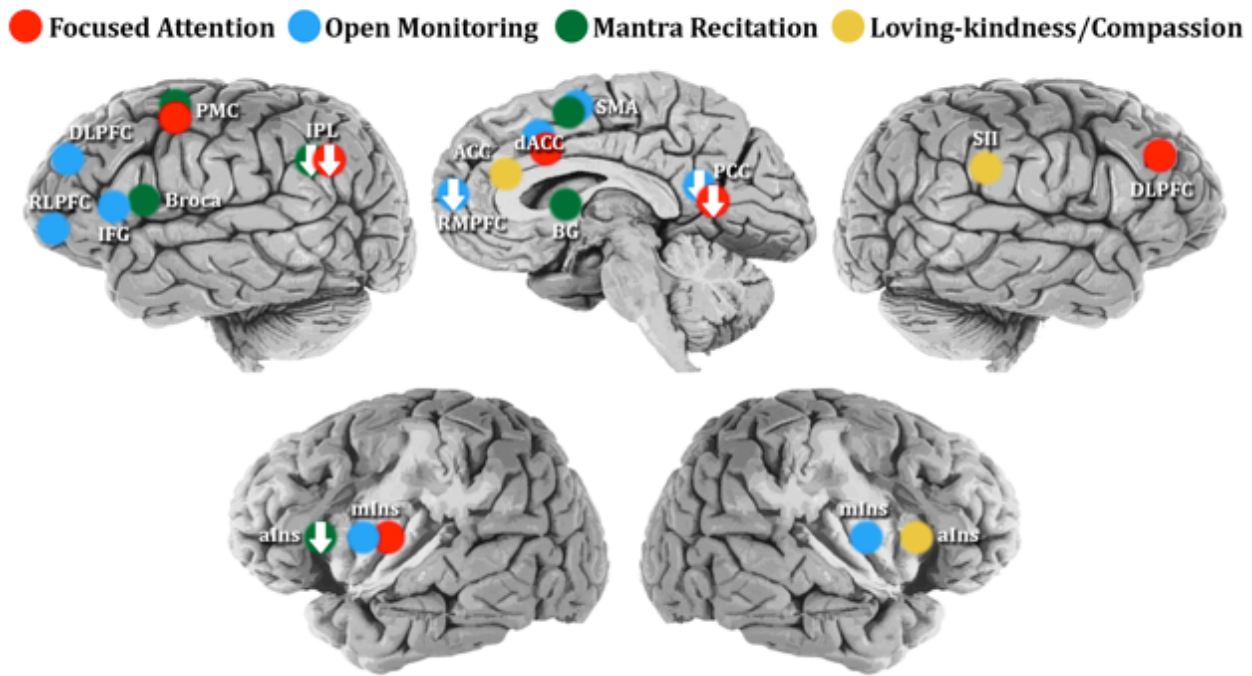

Filled circles indicate activations; circles with downward arrows indicate deactivations. All labels are approximate and intended for illustrative purposes only. Medial meta-analytic clusters are arbitrarily represented on a right hemisphere view merely for simplicity; lateral and insular clusters are represented in the correct hemisphere. For the specific location of each cluster, see Tables 3-6 and Supplementary Figures 1-4.

*5.2 Functional neuroanatomy of focused attention meditation*

**5.2.1 Activations.** Focused attention meditation was associated with activations in regions associated with cognitive control and self-reflection. Directly in line with others' predictions (Lutz et al., 2008b), we found activations in the premotor cortex extending into the posterior dorsolateral prefrontal cortex, as well as in the dorsal anterior cingulate cortex. We also observed a slightly sub-threshold (88 m$^3$) cluster in the dorsolateral prefrontal cortex (BA 8/9). These regions are frequently activated in studies of cognitive control that require monitoring performance, and the voluntary regulation of attention and behavior, such that task-relevant actions are selected (Carter et al., 1998, Vincent et al., 2008, Dixon and Christoff, 2012, Dixon et al., 2014a). The recruitment of such executive brain areas is consistent with a contemplative practice consisting largely of effortful, sustained attention with a range of regulation demands. For more details, see the Supplementary Discussion.

**5.2.2 Deactivations.** Consistent deactivations were observed in two major hubs of the default mode network: the posterior cingulate cortex (BA 23/31) and the posterior inferior parietal lobule (BA 39) (Buckner et al., 2008). These regions have well-established roles in mind-wandering (Fox et al., 2015), and in particular, episodic memory retrieval, simulation of future events, and conceptual/semantic processing. Deactivations in these regions suggest that focused attention meditation may diminish spontaneous thoughts regarding past and future events, as well as their conceptual elaboration. Although many regions beyond the default mode network are implicated in mind-wandering and related spontaneous thought processes, the default mode network nonetheless plays an essential role (Fox et al., 2015). For more details, see the Supplementary Discussion.

*5.3 Functional neuroanatomy of mantra recitation meditation*

**5.3.1 Activations.** The most salient activations for mantra recitation meditation were in the motor control network, including Broca's area, the premotor and supplementary motor cortices (lateral and medial BA 6), as well as the putamen within the basal ganglia. Consistent with the goals of mantra meditation, these regions may be involved in the process of internally generating and staying focused on a phrase within one's mind (or recited aloud).

Particularly interesting was a small cluster in Broca's area (BA 44), extending posteriorly into the face/tongue area of premotor cortex (BA 6). This area is well-known from the neuropsychological lesion work of Pierre Paul Broca to be critically involved in the motor components of speech production (Broca, 1861). Its consistent recruitment during mantra recitation requires little comment.

Large activations were also observed in various areas of the basal ganglia, including bilateral clusters in both the putamen and globus pallidus. Although these areas play a clear role in 'reward' (Schultz et al., 2000), they are most strongly implicated in highly-practiced motor movements and habit formation in general (Yin and Knowlton, 2006, Ashby et al., 2010). In a broader sense, the basal ganglia are thought to aid in the inhibition of undesired movements, and in facilitating the smooth execution of those movements that *are* desired, via tight reciprocal connections with premotor areas (Chakravarthy et al., 2010, Stocco et al., 2010). These latter functions are all consistent with a meditation practice that primarily involves the deliberate execution of a highly-practiced and highly-specific sequence of motor behavior (whether imagined in detail or recited aloud). For more details, see the Supplementary Discussion.

**5.3.2 Deactivations.** An interesting pattern of deactivations was observed, suggesting decreased processing of external sensory (particularly somatosensory) inputs. Deactivations were observed bilaterally in the anterior insula, and at a sub-threshold level in the thalamus, left primary somatosensory cortex (postcentral gyrus; BA 2), and left inferior parietal lobule (BA 40). All of these areas are centrally involved in processing and awareness of somatosensory and interoceptive information (Penfield and Boldrey, 1937, Bonda et al., 1995, Craig, 2004, 2009). Deactivations were also observed in primary auditory cortex (BA 41) and in the posterior area of the superior temporal gyrus (BA 22), both well-known to be involved in the processing and comprehension of sounds and language (Galaburda and Sanides, 1980, Demonet et al., 1992).

One of the reasons mantra is employed is to heighten the effectiveness of focused attention meditation. The implication is that by focusing upon a *single* salient percept (the mantra), attention is more easily sustained and distractors more easily ignored than without the mantra (Goenka, 2000). The observed pattern of deactivations is consistent with the idea that mantra aids in maintaining a sharp focus of attention, to the exclusion of incoming sensory stimuli and awareness of one's own body. Deactivations in auditory and language-comprehension areas are also intriguing in that the mantra is intended as a tool for focusing attention, but is not usually meant to be understood semantically.

*5.4 Functional neuroanatomy of open monitoring meditation*

**5.4.1 Activations.** Open monitoring meditation was associated with activation in regions involved in the voluntary regulation of thought and action (inferior frontal gyrus, posterior dorsolateral prefrontal cortex/pre-motor cortex, and dorsal anterior cingulate cortex/pre-supplementary motor area), as well as activation in the insula, a region that plays a key role in interoceptive processing. As all of these areas were recruited by multiple meditation techniques, they are discussed in more detail below in Section 5.6 on 'Convergent findings.'

Smaller clusters that did not exceed our threshold were observed in left rostrolateral prefrontal cortex (BA 10) and left mid-dorsolateral prefrontal cortex (BA 9/46). Together with the dorsal anterior cingulate cortex, these areas constitute key parts of a 'frontoparietal control' network (Vincent et al., 2008, Spreng et al., 2010) generally involved in cognitive control, that is, coordinating and monitoring attention to both internal and external channels of information (Spreng, 2012, Dixon et al., 2014b). Given that open monitoring meditation typically involves a broad, non-judgmental attention to any and all mental content (including information from the outside world, from within the body, and also internal channels of thought and memories), it is congruous that this practice recruits numerous areas from the frontoparietal control network.

**5.4.2 Deactivations.** Only a single significant deactivation was observed, in the right thalamus, and a sizeable (albeit sub-threshold) deactivation was also observed in the left thalamus. Interestingly, in the supplemental meta-analysis (which excluded practitioners with only brief training in this practice), these findings were even more salient (Table S5). The thalamus is the key relay center for most incoming sensory information, and also functions in 'sensory gating' (i.e., the selective filtering out of certain sensory signals). Interestingly, increased intensity of sensory gating appears to be correlated with *increased* activity in the thalamus (LaBerge, 1990, LaBerge and Buchsbaum, 1990, Banich, 2004). Therefore, deactivations here suggest decreased sensory gating, an interpretation that aligns well with the goal of being open and receptive to sensory stimuli. For more details, see the Supplementary Discussion.

*5.5 Functional neuroanatomy of loving-kindness and compassion meditation*

**5.5.1 Activations.** Loving-kindness and compassion meditation involve cultivating positive emotions, such as joy and compassion, as well as prosocial skills and tendencies, such as empathic concern and altruistic behavior. They focus on active 'mentalizing' in the form of taking the perspective of others and imagining their emotional experiences. Mounting evidence from behavioral studies suggests that these practices can be effective, and that even brief training in compassion meditation can increase empathy for, and prosocial behavior toward, others (Leiberg et al., 2011, Condon et al., 2013, Kang et al., 2013, Weng et al., 2013, Lim et al., 2015). One study even found that compassion meditation-driven changes in brain activation specifically predicted subsequent compassionate behavior toward an unknown victim (Weng et al., 2013).

Only three significant meta-analytic clusters were observed, probably due to the low power of this analysis, which involved the fewest studies ($N$ = 5 experiments; see Table 1). One large cluster was observed in the right somatosensory cortices (BA 2), extending into the inferior parietal lobule (BA 40). A similarly located (albeit statistically non-significant) cluster was observed on the postcentral gyrus of the *left* hemisphere, centered on BA 43 and extending into BA 3. These parts of parietal lobe play a well-known role in somatosensory processing and in creating a unified sense of the body (Penfield and Boldrey, 1937, Penfield and Rasmussen, 1950), but both regions have also been implicated in empathy in a recent meta-analysis (Lamm et al., 2011) and comprehensive review (Bernhardt and Singer, 2012). Both areas are also consistently implicated in the perception of pain (Grant et al., 2011, Grant, 2014), but this function may overlap with their role in empathy, as part of the proposed role of these areas in empathy may be in specifically empathizing with the pain of others (Lamm et al., 2011).

Another large cluster centered on the right anterior insula, which has previously been implicated in quantitative meta-analyses (Lamm et al., 2011) of the neural correlates of empathy and theory-of-mind (i.e., mentalizing). Because insula activation was observed across multiple meditation categories, its possible role(s) in loving-kindness meditation is discussed further in the section on convergent findings (section 5.6). For more details, see the Supplementary Discussion.

**5.5.2 Deactivations.** No significant deactivations were observed; for more details, see the Supplementary Discussion.

*5.6 Convergent findings across meditation categories*

Relatively few regions showed significant clusters across multiple meditation categories, supporting the central hypothesis of this paper that distinct meditation styles recruit largely non-overlapping neural networks. Only a single area (insular cortex) was recruited by *all four* meditation categories. Some additional consistencies across meditation types were observed, however; each area is considered in turn below.

**5.6.1 Insular cortex.** Meta-analytic activations were observed in insular cortex in all four meditation categories explored here. For focused attention, a small (non-significant) cluster was observed in the left anterior insula, as well as a somewhat larger (also non-significant) cluster in the left mid-insula. For mantra recitation, a very small (non-significant) cluster was observed in the left mid-insula/claustrum. For open monitoring meditation, a large (significant) cluster was observed in the left anterior/mid-insula, as well as two smaller (non-significant) clusters in the right mid- and posterior insula. For loving-kindness meditation, a large (significant) cluster was observed in the right anterior insula. The insula therefore appears to be reliably recruited in two categories of meditation we examined, and shows evidence for recruitment in the two other styles.

These findings suggest an important role for the insula across numerous meditative practices. They are also consistent with our recent meta-analysis of morphometric neuroimaging of meditation, which collapsed across many meditation styles and found that differences in insula were the most consistent structural alterations (Fox et al., 2014). Although the insula has many roles, probably the most consistently replicated and thoroughly investigated is its involvement in interoception (i.e., monitoring and conscious awareness of internal body states, such as respiration or heart rate) (Craig, 2004, Critchley et al., 2004, Craig, 2009, Farb et al., 2013) as well as empathy (Lamm et al., 2011) and possibly even metacognition (Fleming and Dolan, 2012). Given that virtually all forms of meditation investigated here involve some degree of monitoring of the body, some awareness of respiration, and basic metacognitive monitoring, universal recruitment of insular cortex is not unreasonable.

Insular cortex is not functionally homogenous, however: there is evidence that the mid and posterior insula serve as 'primary' interoceptive cortex, receiving and processing direct interoceptive inputs from the thalamus (Flynn, 1999, Hua et al., 2005, Frot et al., 2007), whereas more anterior parts of the insula are preferentially involved in conscious awareness of interoceptive signals and the integration of this information with one's higher-order goals and emotional states (Craig, 2004, Critchley et al., 2004, Critchley, 2005, Craig, 2009). In this light it is interesting that anterior insula was recruited in all techniques except mantra recitation – which is the technique least focused on conscious awareness of the body, and most geared toward detaching from immediate sensory stimuli. This interpretation is speculative, of course;

for more details, see the Supplementary Discussion.

### 5.6.2 Premotor cortex and supplementary motor area (lateral/medial BA 6).

Three categories of meditation (the exception being loving-kindness/compassion) showed large meta-analytic clusters in posterior dorsolateral prefrontal cortex/premotor and supplementary motor cortices (BA 6). Although these areas are traditionally considered 'motor' regions, a wealth of recent work has demonstrated that they also play a role in many cognitive functions (e.g., (Hanakawa et al., 2002, Tanaka et al., 2005). Specifically, it has been proposed that these areas could play a role analogous to their motor function of manipulating objects in the physical world – namely, the mental manipulation of psychological content and memory (Hanakawa et al., 2002). Functional neuroimaging studies have shown activation in these areas during a variety of complex mental tasks, including working memory (Fiez et al., 1996, Owen and Evans, 1996), attentional control (Hopfinger et al., 2000, Boussaoud, 2001), mental imagery (Mellet et al., 1996), and conceptual reasoning (Rao et al., 1997). Given the apparent role of these regions in diverse higher-order cognitive functions and attention regulation – the core of all meditation techniques – their recruitment across numerous styles of meditation appears reasonable. Speculatively, then, these regions may interact with other brain areas more specifically tied to particular meditation techniques, assisting in the intentional regulation of attention and other cognitive processes specific to the practice at hand.

**5.6.3 Dorsal anterior/mid cingulate cortex.** In focused attention and open monitoring meditation, we observed significant activation clusters that bordered both the supplementary motor areas (discussed in the previous section) and the dorsal anterior cingulate cortex. Congruent with these findings, consistent structural alterations in dorsal anterior cingulate cortex were observed in our recent meta-analysis of morphometric neuroimaging of meditation (Fox et al., 2014). Despite numerous functional roles, the dorsal anterior cingulate is clearly a key player in the brain's ability to regulate attention and emotion, as well as monitor performance (Bush et al., 2000, Posner et al., 2007). Activations here across the two styles of meditation that most explicitly involve such regulation and monitoring is therefore sensible. For further discussion, see the individual sections on focused attention and open monitoring.

**5.6.4 Frontopolar cortex/rostrolateral prefrontal cortex.** We observed a fairly large but sub-threshold cluster of activation in the left frontopolar cortex (BA 10) for open monitoring meditation. For loving-kindness/compassion meditation, we observed a small (non-significant) cluster in the superior portion of left frontopolar cortex (BA 10), extending somewhat into the dorsolateral prefrontal cortex (BA 9). Although neither of these results exceeded our stringent significance thresholds, nonetheless these findings are in line with our recent meta-analysis of morphometric (i.e., structural-anatomical) neuroimaging studies of meditation (Fox et al., 2014), and therefore warrant a brief discussion.

In our morphometric meta-analysis, we reported anatomical increases (e.g., increased cortical thickness; Lazar et al., 2005) in this area across three fairly divergent schools of practice (Fox et al., 2014). This region is known to play a key role in meta-awareness and metacognitive capacity generally (e.g., (Fleming et al., 2010, McCaig et al., 2011, Fleming and Dolan, 2012, Fleming et al., 2012, Fox and Christoff, 2015)), but it also appears to be involved in the *evaluation* of self-generated information (Christoff et al., 2003) and in the processing of complex and abstract information generally (Christoff et al., 2001, Christoff et al., 2009). Frontopolar cortex is also a key node of the frontoparietal control network (Vincent et al., 2008) and may play an important role in switching attention between external and internal channels (Ramnani and Owen, 2004, Burgess et al., 2007). The specific functions of the frontopolar cortex in various forms of meditation remain an open question.

*5.7 Other meditation practices: Further evidence for dissociability at the neural level*

In the Introduction, we noted that there are several other major categories of meditation wherein research remains too sparse for inclusion in a meta-analysis (Section 1.3). These categories include *visualization* (generating and sustaining complex visual imagery)*, non-dual awareness* (eroding the boundary between subject and object)*,* and *pratyahara* (dampening of sensory inputs). Neural correlates of these practices have been reported in isolated studies, and interestingly, provide further support for the hypothesis that distinct psychological ($\Psi$) practices are likewise dissociable at the neurobiological ($\Phi$) level (Cacioppo and Tassinary, 1990).

For instance, a single study has examined the neural correlates of *visualization* meditation (Lou et al., 1999). In this study, practitioners were guided through "visual imagination of a summer landscape with forests, streams, and meadows with cattle" (Lou et al., 1999; p. 100). Although this particular imagined *content* is not typical of visualization meditation practices (which tend to focus on more explicitly religious themes), nonetheless the results are telling. Visualization meditation resulted in significant recruitment (among other areas) of the fusiform gyrus (definitively linked to higher-order visual processing; (McCarthy et al., 1997, McCandliss et al., 2003), the parahippocampal gyrus (linked to visual 'place' perception; (Epstein et al., 1999), and nearly universal recruitment of medial and lateral occipital areas (Lou et al., 1999). Intriguingly, primary visual cortex (V1), which is not required for the creation of visual imagery and visual dreaming (Solms, 1997), was essentially the only occipital area not strongly recruited (Lou et al., 1999). These results, although drawn from only a single study, support the idea that practitioners were successfully engaging in such a visualization practice. More importantly, for our purposes, they are utterly distinct from any of the broad patterns of neural activity reported in this meta-analysis, further supporting the contention that differing mental practices have dissociable and identifiable neural correlates.

A case study involving a single highly-experienced practitioner engaging in a *pratyahara* (sense-withdrawal) practice is similarly informative (Kakigi et al., 2005). Compared to painful stimulation when not engaging in meditation, which led to recruitment of all the regions commonly implicated in pain processing, during meditation there was a marked deactivation of these classic pain areas, including the thalamus, secondary somatosensory cortex, insula, and cingulate cortex. Again, none of the four practice types meta-analyzed here bears much resemblance to this pattern (and open monitoring seems almost antithetical to it – in line with the methods and motivations underlying each practice).

Finally, a single study investigating *non-dual awareness* in highly experienced Tibetan Buddhist practitioners reported evidence that this practice decreased the anti-correlation between activity in 'extrinsic' and 'intrinsic' networks of the brain, which are argued to mediate attention to the external environment and internal states, respectively (Josipovic et al., 2011). The authors created numerous regions-of-interest throughout the brain corresponding to an 'extrinsic' network for attention to external stimuli, and an 'intrinsic' network thought to be involved in internal streams of thought (which roughly parallels the default mode network). The extrinsic and intrinsic networks are thought to show anti-correlated patterns of functional connectivity during either external tasks or attention to internal streams of thought (e.g., during mind-wandering) (Fox et al., 2005). The authors examined the extent to which this anti-correlated functional connectivity might be attenuated by non-dual awareness meditation, which broadly aims at reducing the sense of a division between external and internal. During non-dual awareness meditation, the anti-correlation between these two networks was markedly attenuated compared to either focused attention meditation or passive fixation. The authors interpreted these results as evidence of the effectiveness of non-dual awareness practice at dissolving the boundary between internal and external experiences (i.e., subject-object duality; Josipovic et al., 2011). In line with our central hypothesis and meta-analytic results, the authors also suggested that their results indicated that non-dual awareness practice is differentiable at the neural level from other meditations, such as focused attention and open monitoring meditation (Josipovic et al., 2011).

Although the evidence regarding these other forms of meditation remains tentative, it seems clear that further research is warranted into various other techniques that have not yet been intensively investigated. Based on our meta-analytic findings, as well as the studies just reviewed, we suspect that future work will discover and refine dissociable neural correlates associated with these various psychologically unique meditation practices.

*5.8 Are contemplative practices dissociable by electrophysiological and neurochemical measures?*

Virtually all of the meta-analytic results and discussion in this review have relied on brain signals indicative of cerebral blood flow (PET) or blood oxygenation (fMRI), but there are numerous other ways of characterizing brain function, most notably via electrophysiological activity (via electroencephalography, for instance; EEG) and at the neurochemical level (i.e., the tonal and phasic rates of release of various neurotransmitters across different areas of the brain). For more details on how these possibilities have been applied in meditation research, see the Supplementary Discussion.

*5.9 Do functional neural effects associated with meditation have practical significance?*

Our effect size meta-analyses found that the four primary categories of meditation examined were all associated with approximately 'medium' effect sizes (mean Cohen's *d* = 0.60 for activations and -0.74 for deactivations; for details see Tables 7 and 8). Although we find these results both reasonable (in magnitude) and encouraging, we reiterate that the 'meaningfulness' of effect sizes is ultimately arbitrary even where their use is highly developed (e.g., in the behavioral, social, and medical sciences; Cohen, 1992). This is all the more true in neuroimaging, where their calculation and interpretation are poorly understood (see Supplementary Methods). Nonetheless, some general observations can be made.

First, positive and negative effect sizes (representing activations and deactivations associated with meditation, respectively) are similar in magnitude (Tables 7 and 8; Fig. 9). This is in line with our meta-analytic neuroimaging findings (where brain deactivations were nearly as common and significant as activations) and supports the idea that the successful execution of a particular meditation may depend as much on which brain areas are *disengaged* as on which are actively recruited.

Second, the approximately 'medium' magnitude of the effects suggests that the functional neural effects of meditation result in roughly a one-half standard deviation increase or decrease in brain activity, relative to various other baseline conditions, or novices attempting to execute the same practice as experts.

Third, differing meditation categories showed largely similar mean effect sizes – roughly one half to two-thirds of a standard deviation (Cohen's $d$ ~0.50 – 0.67; see Table 8). The exception was mantra recitation, which appeared to result in larger effects for both activations and deactivations (Cohen's $d$ > 1.0). This apparently large difference is highly questionable, however: first, these 'mean' (+) and (–) effect sizes across studies are in fact simply mean effect sizes for a *single* study each, as only two studies contributed to the effect size meta-analysis for mantra recitation (see Tables 7 and 8). Moreover, the two studies that contributed (Davanger et al., 2010, Kalyani et al., 2011) have among the smallest sample sizes of any studies we investigated here (see Table 7). This makes these reports highly susceptible to the inflationary biasing of effect sizes we discussed at length in the Supplementary Methods. We therefore conclude that it would be highly premature to argue that mantra meditation results in larger effects than other forms of meditation practice. A more plausible interpretation is that inflationary bias was larger in these studies, and the true effect size of mantra recitation is in fact comparable to the other practices examined here.

Fourth, a funnel plot (scatterplot of effect size vs. sample size; Fig. 6) showed a conspicuous dearth of null results or small effect sizes (even after our multiple deflationary adjustments of originally reported effect sizes), for both 'positive' effect sizes (brain activations) and 'negative' effect sizes (brain deactivations). The simplest interpretation of these results is that fairly serious publication bias (i.e. non-publication of null and 'non-significant' findings) afflicts this field. Even if meditation always and truly results in significant effects on brain activation or deactivation, given that 78 functional neuroimaging studies had been published at the time of writing, we would expect at least a few null results to have been obtained and reported simply by chance alone. An alternative (albeit related) explanation is that smaller effect sizes were not observed precisely because of the inflationary bias inherent in neuroimaging reporting of effect sizes (see section 2.5). Although this possibility presents at least as large a problem as does publication bias, it shifts the emphasis from the limitations of the current scientific publication model (which places undue emphasis on 'positive' findings) to the limitations of neuroimaging data analysis and reporting practices. In either case, the lack of reported small or null effects in this literature (even after our severe deflationary procedures) suggests bias at some stage of study design, analysis, or reporting.

That said, for both negative and positive effect sizes, the expected funnel plot pattern was generally observed: that is, with increasing sample size, the variability among effect sizes reported decreased drastically; and notably, the most extreme effect sizes were always reported in studies with the smallest sample sizes. This suggests, encouragingly, that despite potential publication bias or inflationary bias due to neuroimaging analysis methods, nonetheless studies with larger samples tend to converge on similar and more reasonable (medium) effect sizes. Although such a conclusion is tentative, the results to date (Fig. 6) suggest that a sample size of approximately $n = 25$ is sufficient to reliably produce effect sizes that accord with those reported in studies with much larger samples (up to $n = 46$).

*5.10 Meditation, the brain, and clinical disorders*

Although the central aim of this meta-analysis was to better delineate the neural correlates of meditation in healthy practitioners, the potential effects of meditation on various physical and psychological clinical conditions is of enormous interest (Gotink et al., 2015) and therefore warrants a brief discussion. To our knowledge, only five studies to date (Table S1) have examined the relationship between meditation, brain activity, and clinical disorders of some kind. One study investigated social anxiety disorder (Goldin and Gross, 2010) and another bipolar disorder (Ives-Deliperi et al., 2013); one examined patients with breast cancer (Monti et al., 2012); one involved chronic tobacco smokers during a period of abstinence from smoking (Westbrook et al., 2013); and one recruited patients with mild cognitive impairment (Wells et al., 2013). The heterogeneity of these investigations precludes any quantitative meta-analytic procedure, but nonetheless these preliminary explorations are a very important step toward examining the potential of meditation to ameliorate such conditions, as well as understanding the cognitive and neural mechanisms by which these improvements might take place. Considering the fairly ample evidence from psychological research that meditation can have salutary effects on affective characteristics (Sedlmeier et al., 2012, Goyal et al., 2014), further investigations into its potential benefits for clinical conditions, and the brain mechanisms related to these benefits, are of paramount importance.

*5.11 Meta-analytic limitations*

First, it should be reiterated that the four categories employed in our synthesis combined meditation practices which, although similar, are not necessarily identical. Moreover, there are several other major categories of meditation that could not be quantitatively examined here (e.g., non-dual awareness and sensory-withdrawal practices), due to the paucity of empirical data to date (see Section 1.3).

Second, despite inclusion of all eligible studies to date, the present analysis relies on relatively few research reports ($n = 25$; see Table 1). We also combined results from long-term practitioners with those of novices undergoing short-term training in an effort to include all available data (see Table 2). This may be problematic because, as noted earlier, the diverse goals of meditation practices can take considerable time to be realized. It might therefore be expected that the distinctions between brain activation patterns driven by particular forms of meditation may be even sharper among expert vs. novice practitioners, or that brain activity in novices might otherwise influence or drive the effects observed here. To mitigate these concerns, however, we conducted a series of four supplemental meta-analyses where we excluded studies of short-term meditation training, and found little evidence that this was the case. Meta-analytic results for each meditation category employing *only* long-term practitioners were negligibly different from our main meta-analyses, except in the case of open monitoring meditation. In the case of open monitoring meditation, the general pattern of results was also similar, but the precise locations and significance levels of various clusters were different (for details, compare Table 5 with Table S4).

Third, study methods and practitioner samples were highly heterogeneous. Both meditation and baseline conditions differed considerably across studies (even within a single meditation category), as did meditation experience level (even among expert practitioners).

Fourth, numerous and serious difficulties attend the calculation and interpretation of effect sizes in neuroimaging data. Our results in this sphere, and their bearing on the question of the 'practical significance' of meditation's effects on brain function, need to be viewed in light of the many caveats and limitations we have discussed at length in the Supplementary Methods.

Future work may be better able to mitigate many of these limitations by directly examining multiple forms of meditation within the same practitioners and/or within the same study, and using more homogenous baseline/control conditions (e.g., Brewer et al., 2011). Such an approach would control for many of the potential confounds that plague comparison across different studies, conducted by different researchers, investigating different samples of practitioners. Taken together, the present results should be interpreted with caution: a clearer understanding of the major categories of meditation, as well as their subtler subdivisions, awaits further research undertaken with more directly comparable methods, baselines, and populations.

*5.12 Other methodological challenges in the neuroimaging of meditation*

Other major concerns have less to do with meta-analytics or statistical issues, but rather relate specifically to the investigation of meditation practitioners. Several important themes are outlined here.

**5.12.1 Lifestyle variables.** The first concern is the general and unknown effect of *lifestyle*. Highly-experienced meditation practitioners are likely to have more meager (calorie-restricted) and more vegetarian diets; are less likely to have children or long-term regular employment; and probably sleep less than the average person (Britton et al., 2014). These are just a few major examples; an array of other factors might be relevant. Different meditation traditions emphasize such factors to varying degrees, and it is certainly conceivable that major differences in diet, stress, and sleeping patterns could have effects on brain structure and function. Control subjects in the studies meta-analyzed here, however, typically only control for simple variables like age, sex, and sometimes handedness, and are therefore unable to account for these lifestyle variables. Until these and other factors are carefully controlled for, the potentially confounding effects of non-meditation-specific aspects of the contemplative lifestyle cannot be ruled out or fully understood (Fox et al., 2014).

**5.12.2 Pre-existing differences in brain structure and function.** A second general concern is the possibility of *pre-existing differences* in brain structure or function in advanced meditation practitioners (Fox et al., 2014). Given the rigorous and peculiar demands of spending thousands of hours in solitary, minimally-stimulating environments and unusual (for Western practitioners) and often painful physical postures, it seems highly likely that long-term contemplative practice may preferentially attract persons with a proclivity toward solitude and sustained attention to inner mental life. Support for this hypothesis comes from a recent study that was able to predict subsequent practice of two different forms of meditation (mindfulness and compassion) from baseline brain activity alone (Mascaro et al., 2013). Cross-sectional studies (which form a large part of this meta-analytic dataset) cannot control for such pre-existing differences. On the other hand, studies involving short-term training of novices in pre-post designs cannot examine the potentially subtle effects that long-term and committed practice may have on brain function. Future work will have to strike a balance somewhere between the limitations of these two general approaches, and further studies of pre-existing (and potentially predictive) differences in brain structure and function in meditation practitioners would be highly beneficial (cf. Mascaro et al., 2013).

**5.12.3 'Overshadowing.'** A third and major concern is what we call 'overshadowing' – the possibility that a practitioner's background experience with a variety of meditation techniques might color the neurophysiological results obtained during the brief 'state' investigations of isolated meditation techniques that are typical of the experiments examined here. This issue is discussed further in the Supplementary Discussion.

**5.12.4 Enduring trait differences in brain and cognitive function.** Closely related to the overshadowing effect are unknown *trait* differences. By 'overshadowing' we mean the possibility that accumulated trait differences might overshadow (or interact with) brain activations and deactivations directly associated with a brief, induced *state* of meditation in neuroimaging investigations. But although these putative trait differences are a potential confounding factor for our purposes here (investigating meditation *state* effects), enduring trait changes are an explicit goal of most meditation traditions, and are a worthy object of study in their own right. Here, we chose to exclude such studies from our meta-analysis (criterion iv; see Methods section 2.1.2), mainly because of the enormous methodological challenges involved in synthesizing across so many disparate tasks and contemplative traditions. Although our meta-analysis was also affected by differing baseline/comparison conditions, and potentially also an 'overshadowing' effect, nonetheless the meditation states investigated were constant and consistent for each analysis (focused attention, mantra meditation, etc.). The major challenge facing any effort to synthesize *trait* effects associated with meditation practice is that even this limited homogeneity (constant meditation states, despite differing samples and comparison conditions) is absent. Not only is there great heterogeneity among the tasks on which meditation-related traits are being investigated (e.g., emotion regulation, perceptual discrimination, etc.), no consistent meditation state is being induced, and the practitioners employed will invariably possess heterogeneous contemplative backgrounds with respect to the proportion, and absolute amount, of time spent in various different practices. One way to begin to examine such trait differences, which sidesteps these problems, would be to compile data on long-term practitioners during the 'resting' state, for instance measuring changes in functional connectivity at baseline (in the absence of any task or stimuli).

**5.12.5 Integration of functional neuroimaging methods with behavioral measures.** As we discussed in our previous meta-analysis of morphometric neuroimaging studies of meditation (Fox et al., 2014), alterations in brain structure and function are of little value (beyond basic scientific interest) if they are not accompanied by changes in behavior or cognitive-emotional experience and wellbeing (Fox et al., in press). We discuss this issue and some of the findings we reviewed related to brain-behavior correlations in the Supplementary Discussion.

*5.13 Summary and conclusions*

The present meta-analysis is the first to report dissociable patterns of brain activation and deactivation underlying four major categories of meditation (focused attention, mantra recitation, open monitoring, and loving-kindness/compassion practices), and provides suggestive evidence for the dissociability of three additional meditation types (visualization, sense-withdrawal, and non-dual awareness practices). Our synthesis provides preliminary evidence for largely dissociable neural substrates which, broadly speaking, are consistent with the methods, aims, and putative psychological results of the practices examined. Collectively, these results suggest that while several regions may be equally involved in many forms of contemplative practice (Sperduti et al., 2012), differences in neural activity greatly outnumber similarities. *Commonality across meditation categories is the exception rather than the rule.* Importantly, deactivations almost always accompany activations (though they are usually fewer in number, less substantial in terms of cluster size, and less consistent across studies in terms of peak likelihood value). This suggests that for the effective execution of various meditation practices, disengaging particular brain regions may be nearly as important as engaging others.

After careful corrections and adjustments for suspected inflationary biases, effect size meta-analysis suggests that meditation practices tend to yield approximately medium-sized effects. These results suggest that meditation's relationship with brain function may be practically significant, but there appears to be a long way to go in the statistical theory of neuroimaging before these findings can be interpreted with confidence. Moreover, it is imperative that neuroimaging investigations of meditation begin to rigorously relate findings to behavioral and self-reported outcomes, in order to understand how (and to what extent) these changes in brain activity are related to actual cognitive-affective change and benefits in practitioners.

Many have understandably viewed the nascent neuroscience of meditation with skepticism (Andresen, 2000, Horgan, 2004), but recent years have seen an increasing number of high-quality, controlled studies that are suitable for inclusion in meta-analyses and that can advance our cumulative knowledge of the neural basis of various meditation practices (Tang et al., 2015). With nearly a hundred functional neuroimaging studies of meditation now reported, we can conclude with some confidence that different practices show relatively distinct patterns of brain activity, and that the magnitude of associated effects on brain function may have some practical significance. The only totally incontrovertible conclusion, however, is that much work remains to be done to confirm and build upon these initial findings.

**Acknowledgments**

We thank Dr. Antoine Lutz and Dr. David M. Fresco for very helpful discussions of the methods and results of this meta-analysis. We also thank Dr. Norman Farb for providing further details from previously unpublished data that contributed to the effect size meta-analysis. This work was supported in part by Natural Sciences and Engineering Research Council (NSERC) Vanier Canada Graduate Scholarships awarded to K.C.R.F. and M.L.; NSERC Canada Graduate Scholarships awarded to M.L.D. and M.E.; a University of British Columbia Affiliated Doctoral Fellowship awarded to J.L.F.; and NSERC and Canadian Institute of Health Research (CIHR) grants awarded to K.C.

# • SUPPLEMENTARY MATERIAL •

**Functional neuroanatomy of meditation: A review and meta-analysis of 78 functional neuroimaging investigations**

Kieran C. R. Fox[a,*], Matthew L. Dixon[a], Savannah Nijeboer[a], Manesh Girn[a], James L. Floman[b], Michael Lifshitz[c], Melissa Ellamil[d], Peter Sedlmeier[e], and Kalina Christoff[a,e]

[a] Department of Psychology, University of British Columbia, 2136 West Mall, Vancouver, B.C., V6T 1Z4 Canada

[b] Department of Educational and Counselling Psychology, and Special Education, University of British Columbia, 2125 Main Mall, Vancouver, B.C., V6T 1Z4

[c] Integrated Program in Neuroscience, McGill University, 3775 University St., Montreal, QC, H3A 2B4

[d] Neuroanatomy and Connectivity Research Group, Max Planck Institute for Human Cognitive and Brain Sciences, Stephanstrasse 1a, Leipzig, Germany 04103.

[e] Institut für Psychologie, Technische Universität Chemnitz, 43 Wilhelm-Raabe Street, Chemnitz, Germany

[e] Brain Research Centre, University of British Columbia, 2211 Wesbrook Mall, Vancouver, B.C., V6T 2B5 Canada

[*] To whom correspondence may be addressed (at address [a] above). Telephone: 1-778-968-3334; Fax: 1-604-822-6923; E-mail: kfox@psych.ubc.ca

**Supplementary Methods**

*Further details regarding excluded studies*

About two-thirds (52 of 78) of all studies did not meet with one or more of our inclusion criteria (Table S1). This fact should not be considered an indication of the poor quality of these studies; only a small minority was excluded because of methodological problems (details in Table S1). Many studies were excluded simply because their design or analyses were not amenable to our meta-analytic methods and questions, with criteria (i) and (iv) accounting for the vast majority of exclusions (Table S1). For instance, many studies were excluded because they did not have participants actually engage in meditation practice in the scanner: these studies were interested instead in the effect of prior meditation training on the neural correlates of various *other* tasks during scanning, whereas *our interest was in neural correlates of the meditation practices themselves.*

Mantra recitation bears some resemblance to prayer, but differs in several important ways. For example, mantra recitation is typically impersonal (the practitioner may not even know the meaning of the words being recited, which are often in an ancient and unfamiliar language, such as Sanskrit), whereas prayer often involves a personal request with meaningful content. Mantra recitation is also often recited in a manner that deliberately matches the recitation to the rhythm of inhalation and exhalation, thus drawing attention toward interoceptive information, whereas prayer does not typically have this component. Therefore, the handful of studies that have investigated various forms of Judeo-Christian prayer (e.g., (Schjoedt et al., 2009, Schjoedt et al., 2010) were not included in our analyses here.

Several other studies pursued the question of meditation training's relationship to various clinical conditions and the brain. Despite the high value of such studies, their inclusion would present a serious confound to our meta-analytic results and so these studies too were excluded. On the other hand, many studies (16 of 78) were excluded because they did not report full tables of voxel coordinates of peak difference for the contrast of interest (meditation > some other condition). Although some of these studies might have been excluded in any case (for a variety of other reasons), nonetheless the failure to report full voxel coordinates represents a serious and unfortunate loss of data for meta-analyses.

*Instructions given in loving-kindness and compassion meditation studies*

Likely the category in our meta-analysis with the greatest potential for variation among studies is the loving-kindness/compassion group. Upon reviewing the instructions given to participants in these six studies, however, we found that overall the similarities greatly outweighed the differences: broadly speaking, all studies had participants generate positive emotions (loving-kindness, well-wishing, compassion, etc.) toward others and/or in a general, non-referential way. Here we reproduce the exact instructions given in each study for the reader's convenience.

• Lou et al. (1999): "...experience of joy and happiness in abstract form (i.e., not related to external events or facts)."

• Lutz et al. (2008): "During the training session, the subject will think about someone he

cares about, such as his parents, sibling or beloved, and will let his mind be invaded by a feeling of altruistic love (wishing well-being) or of compassion (wishing freedom from suffering) toward these persons. After some training the subject will generate such feeling toward all beings and without thinking specifically about someone. While in the scanner, the subject will try to generate this state of loving kindness and compassion."

• Lutz et al. (2009): Same as Lutz et al. (2008).

• Brewer et al. (2011): "Please think of a time when you genuinely wished someone well (pause). Using this feeling as a focus, silently wish all beings well, by repeating a few short phrases of your choosing over and over. For example: May all beings be happy, may all beings be healthy, may all beings be safe from harm."

• Lee et al. (2012): Specific instructions not provided, but described as "very similar" (p. 3) to those used in Lutz et al. (2008) and (2009).

• Weng et al. (2014): "Participants practiced compassion for targets by 1) contemplating and envisioning their suffering and then 2) wishing them freedom from that suffering. They first practiced compassion for a Loved One, such as a friend or family member. They imagined a time their loved one had suffered (e.g., illness, injury, relationship problem), and were instructed to pay attention to the emotions and sensations this evoked. They practiced wishing that the suffering were relieved and repeated the phrases, "May you be free from this suffering. May you have joy and happiness." They also envisioned a golden light that extended from their heart to the loved one, which helped to ease his/her suffering. They were also instructed to pay attention to bodily sensations, particularly around the heart. They repeated this procedure for the Self, a Stranger, and a Difficult Person. The Stranger was someone encountered in daily life but not well known (e.g., a bus driver or someone on the street), and the Difficult Person was someone with whom there was conflict (e.g., coworker, significant other). Participants envisioned hypothetical situations of suffering for the stranger and difficult person (if needed) such as having an illness or experiencing a failure. At the end of the meditation, compassion was extended towards all beings. For each new meditation session, participants could choose to use either the same or different people for each target category (e.g., for the loved one category, use sister one day and use father the next day)."

*Supplementary neuroimaging meta-analyses*

In order to address the potential confound of collapsing across studies investigating long-term practitioners, and those involving novices undergoing short-term training, we executed a series of supplementary meta-analyses where all studies involving short-term meditation training were excluded. This allowed us to ensure that the potentially large differences observed in novices learning meditation for the first time were not driving our meta-analytic results.

Five of the 26 included studies involved analyses of practitioners engaging in short-term meditation training *only* (Farb et al., 2007, Dickenson et al., 2013, Farb et al., 2013b, Weng et al., 2013, Lutz et al., 2014) and were excluded from the supplementary meta-analyses (for details, see Table 2). Another study mixed long-term practitioners and novices in its sample (Kalyani et al., 2011) and did not report results separately for long-term practitioners, and was therefore also excluded. Another study also employed both long-term and short-term practitioners (Taylor et al., 2011), but here the novice practitioners served as the comparison group, and therefore the results reported from this study remained valid for inclusion in the supplementary meta-analyses (the study used a between-group design, comparing brain activity in highly-experienced practitioners to a less experienced group – consistent with the other reports included in the meta-analyses).

Ultimately, six studies were excluded: one each from the focused attention (Dickenson et al., 2013), mantra meditation (Kalyani et al., 2011), and loving-kindness/compassion (Weng et al., 2013) analyses, and three studies from the open monitoring meta-analysis (Farb et al., 2007, Farb et al., 2013b, Lutz et al., 2014). To anticipate our findings, these exclusions generally had only negligible impact on our meta-analytic results, and are therefore reported below in the Supplementary Materials Results.

*Effect size meta-analysis: Discussion of challenges, methods, and equations used*

Should effect sizes even be calculated from neuroimaging data? Intensive debate is ongoing and has not yet provided any definitive answers (Kriegeskorte et al., 2009, Vul et al., 2009, Friston, 2012). There appear to be at least two major recurring concerns. The first is that, typically, only *maximum t*-statistics are reported in neuroimaging analyses. These statistics represent the most extreme peak for the voxel showing the greatest difference among an entire *cluster* of significantly different voxels – the rest of which all have smaller (yet still significant) *t*-statistics. Therefore, the maximum (or 'peak') *t*-statistic is necessarily an overestimate of the average *cluster-wide t*-statistic, which is rarely reported. A straightforward (albeit approximate and preliminary) solution to this problem, which we have employed before (Fox et al., 2014), is presented below (see Section 2.5.6). In general, the approach is to adjust (i.e., deflate) the calculated effect sizes based on *maximum/peak* statistics, so that they are more conservative and more representative of the expected *cluster-wide* values.

A second major concern is that meta-analyses of mean effect sizes should also include all null (non-significant) results. Null results are rarely reported in neuroimaging studies, however. Including all results is important because statistically 'significant' results need not be large in magnitude, particularly if sample size *is* large: with a large enough sample, even very small differences between groups that are trivial for practical purposes can yield small *p*-values that exceed arbitrary-but-conventional significance thresholds (e.g., *p* <.05). Effect size meta-analysis aims to comprehensively include even *non-*significant results for the simple reason that an effect can be large and practically meaningful, but still fail to attain an arbitrary statistical significance threshold due to small sample size, large variability in the data, or a host of other reasons (Schmidt, 1992, Schmidt and Hunter, 1997). By impartially including all reported results (regardless of statistical 'significance') effect size meta-analysis aims to estimate the magnitude of true effects relatively independently of sample size or arbitrarily chosen 'significance' valuations (Schmidt, 1992, Schmidt and Hunter, 1997).

The problem is that standard procedure in neuroimaging studies is to set an arbitrary significance threshold and report only differences that *exceed* this value. Such a practice produces an overwhelming bias toward reporting 'positive' over null results, a broader problem in essentially all fields of research (i.e., the 'file drawer' problem; (Rosenthal, 1979). Effect sizes based solely on results that have attained statistical significance (e.g., *p* < .05) will inevitably *overestimate* the true effect size (Schmidt, 1992).

This appears to be all the more true for 'underpowered' studies with small sample sizes (Kraemer et al., 1998).

Noting that the failure to publish *overall* null findings plagues neuroimaging research as much as other fields, an additional problem is whether to report *within-study* null results. Typical neuroimaging subtraction contrasts involve thousands (or tens of thousands) of independent *t*-tests across the many voxels in the brain. Clearly, reporting all non-significant results would be impractical, and reporting effect sizes for only a small number of voxels may therefore be an "unavoidable" practice in neuroimaging (Friston, 2012). For this same reason, *peak t*-statistics for a given significant cluster are usually the only values reported: unique *t*-statistics for each voxel in a cluster are not usually reported because this could still amount to tens or hundreds of individual results to be listed. Therefore, there seems to be fairly reasonable justification for the practice in neuroimaging of not reporting *within-study* null results (and even significant, *non*-peak results within a cluster).

Unfortunately, this approach (a practical but crude solution to a very difficult problem) introduces an unknown amount of inflationary bias in the mean effect sizes for neuroimaging studies. Monte Carlo simulations suggest that this inflationary bias becomes increasingly large at more stringent significance thresholds (Lane and Dunlap, 1978, Brand et al., 2008) – and unfortunately, highly conservative family-wise error corrections are the norm in neuroimaging studies. Although various cross-validation methods have been suggested to correct for this bias, these methods necessarily employ some of the data for validation and only the remaining portion for estimating effects themselves, and are therefore bound to miss an unknown number of effects (Friston, 2012).

Theoretical arguments have been made, however, that this problem may be less severe in neuroimaging than (for instance) behavioral or clinical investigations. Due to the high cost of neuroimaging, studies almost always involve small sample sizes (typically, $N$ = 10-20). Indeed, for the studies included in the present meta-analysis, samples were as small as just 4 subjects in one repeated-measures design (Davanger et al., 2010), and the mean sample size (including controls) across all included studies was only ~22 subjects (see Table 2). The argument is that smaller sample sizes actually have the benefit of rendering it *less* likely that sample size alone will drive 'significant' effects (Friston et al., 1999, Friston, 2012). But this reasoning only goes so far: stringent significance thresholds for the *t*-tests upon which the effect sizes are based, combined with low statistical power, will nonetheless produce inflations of the reported effects compared to the true effect size, due to an unintentional publication bias toward 'significant' results (Schmidt, 1992, Yarkoni, 2009, Yarkoni et al., 2010, Hupé, 2015). Additionally, underpowered studies are more likely to demonstrate issues with replicability and reliability, presenting a challenge to developing a robust evidence base in any field of research (Cumming, 2012, 2013).

**Calculation of effect sizes (Cohen's *d*).** Wherever possible, we used studies' *t*-statistics of group differences between brain activations of meditators vs. controls, or of meditation practitioners before and after short-term training, to calculate an effect size (Cohen's *d*) for each result, using Equation 1 (Ray and Shadish, 1996):

$$d = t \sqrt{\frac{1}{n_e} + \frac{1}{n_c}} \qquad (1)$$

In this equation, $t$ is the value of the reported peak $t$-statistic, and $n_e$ and $n_c$ are the sample sizes for experimental (meditation) and control groups, respectively (see section 2.5.3 for justification of the use of this equation with repeated-measures designs).

Occasionally, where only $F$-statistics were provided, effect sizes were calculated using the online Practical Meta-Analysis Effect Size Calculator (Lipsey and Wilson, 2001) maintained by the Campbell Collaboration (available at:

http://www.campbellcollaboration.org/escalc/html/EffectSizeCalculator-Home.php).

Studies reporting only $p$-values or $Z$ scores were not included in the effect size meta-analysis because of the inherent difficulty of calculating effect sizes from these statistics without access to original datasets.

**Calculating effect sizes for repeated-measures designs.** Many of the studies examined here involve some form of repeated-measures design, in that they compared brain activation in meditation practitioners with these practitioners' own brain activity during different tasks (e.g., resting and not meditating) or at various time-points (e.g., pre- and post-training in a meditation technique).

Although fairly simple solutions for calculating effect sizes from $t$-statistics derived from repeated-measures designs have been proposed, unfortunately the information required to do so (namely, the correlation between the scores being compared in the $t$-test) is almost never reported (Dunlap et al., 1996). Using Monte Carlo simulations, however, Dunlap and colleagues (1996) have shown that use of the standard effect size formula for between-groups designs only very minimally affects the results, compared to the corrected formula they advocate (their Equation 3). We therefore used the standard formula (our Equation 1) for calculating effect sizes for repeated-measures designs.

**Positive vs. negative mean effect sizes.** After calculating effect sizes for 156 unique results, both positive (+) and negative (–) mean effect sizes were calculated for each of the 22 experiments from 18 studies (Table S2). We did not pool positive and negative effect sizes together because the dependent variables (brain activation or deactivation, respectively) have no intrinsic 'valence' or value. Increased brain activity is not necessarily 'good' and decreased activity not necessarily 'bad' – and vice versa. As a result, pooling positive and negative effect sizes would have had the result that any study reporting both activations and deactivations associated with meditation (of which there were many) would tend toward a mean effect size of zero. This would suggest no practically significant effects, an obviously unwarranted interpretation. The nature of most functional neuroimaging subtraction ($t$-statistic-based) contrasts requires that activations and deactivations *cannot* both be reported in exactly the same area of the brain (although they could be proximate): a study reporting both positive and negative effects in meditation practitioners is not suggesting that the overall effect of meditation approaches zero for a given brain area. Rather, the results suggest that meditation leads to *increased* activation in some parts of the brain, and *decreased* activation in other areas that are *non-overlapping* (whether these are distal or proximal).

**Mean effect size for different meditation categories.** Combining mean effect sizes from each experiment, we further calculated mean effect sizes by meditation category (focused attention vs. mantra recitation, and so on) to investigate whether particular meditation practices have differential effects on brain activation and/or deactivation. In line with the recommendations of Schmidt et al. (2009), we calculated 95% confidence intervals for mean effect sizes, to provide an estimate of their precision, following equations 8 and 10 in Sedlmeier et al. (2012).

**Adjusting for inflation of effect sizes due to reporting of *t*-statistics only from peak voxels.** Standard procedure in neuroimaging research is to report *peak t* or *F* statistics only, which by definition are the *extreme* (maximum) values for a given cluster of significant difference between groups. Mean *t*-statistics for an *entire cluster* of significant differences are rarely reported. The effect sizes we calculated therefore represent the peak, *maximum* effects for each given cluster of results; the actual mean *t*-statistics for entire clusters are necessarily lower. Therefore, our results necessarily overestimate the effect size of the cluster as a whole.

In our recent meta-analysis of morphometric neuroimaging studies of meditation, we attempted to mitigate this inflation problem by examining a single study in which both effect sizes based on *peak t*-statistics and those based on *mean t*-statistics for the entire cluster of significant difference were reported (Fox et al., 2014). This allowed us to approximate the inflationary bias caused by reporting of only peak vs. entire clusters of *t*-statistics, then adjust (deflate) the mean effect sizes from other studies accordingly. We found that average cluster *t*-statistics were about half (~57%) of the value of the peak *t*-statistics for a given cluster (Fox et al., 2014).

Unfortunately, reporting of both peak *and* cluster-wide *t*-statistics is similarly rare in functional neuroimaging studies of meditation. We are aware of only one study involving meditation practitioners that reports both types of results (see Table 1 in (Grant et al., 2011). This study did not meet our inclusion criteria, but this was simply because actual meditation practice (which we sought to meta-analyze) was not engaged in during MRI scanning (see Table S1). Instead, putative trait differences in highly experienced practitioners were examined. That said, given that this study contrasted experienced meditation practitioners with control subjects, it provides a starting point for ascertaining the degree of inflationary bias caused by calculating effect sizes solely from peak, rather than average, *t*-statistics for a given cluster.

Accordingly, although we consider this step absolutely critical to gaining a more accurate assessment of neuroimaging-related effect sizes, our only option was to base our deflationary adjustment on data from a single study. The average *t*-statistics (absolute value, ignoring sign) reported in the study by Grant and colleagues (2011) were 2.952 for cluster-wide results and 4.236 for maximum *t*-statistics (see Table 1 in (Grant et al., 2011). These results suggest that cluster-wide *t*-statistics are approximately 70% of those reported for maximum *t*-statistics (2.95/4.24 = 0.697), roughly in line with our prior efforts to deflate effect sizes in morphometric investigations of meditation (i.e., 57%; (Fox et al., 2014).

Assuming a roughly comparable difference between cluster and peak effect sizes for other studies, we adjusted all our effect sizes accordingly, deflating them to 69.7% of their original value. We readily acknowledge the real and significant limitations of this approach, particularly the uncertainty surrounding the amount of inflation for which adjustment is required. Whether our adjustment (deflation) of effect sizes is too conservative, too lenient, or approximately accurate, remains to be seen, pending development of better understanding of, and better tools for assessing, practical significance in neuroimaging. For present purposes, however, we considered this approach preferable to reporting effect sizes that are *guaranteed* to be overestimates. Despite its limitations, our approach provides a more conservative estimate of practical significance in the functional neuroimaging of meditation than would be provided by unadjusted effect sizes.

**Adjusting for inflation of effect sizes due to selective reporting of statistically significant results.** There have long been serious and effective efforts to account for Type I statistical errors (false positives) in psychological research in general, as well as in neuroimaging research, where massive numbers of multiple comparisons (often thousands of tests) are the norm. However, this sustained effort to control for the rate of Type I errors has been paralleled by a serious failure to control for the rate of Type II errors (false negatives) – that is, the chance of failing to detect a true effect when one is indeed present (Sedlmeier and Gigerenzer, 1989, Schmidt, 1992). Among many other concerns, one important consequence of this situation is that if results are reported only when they exceed a given statistical significance threshold (which is the case in nearly all psychological and neuroscientific research), there will tend to be a bias toward reporting of effect sizes larger than the true effect – and this problem appears to be particularly true of underpowered (small sample) studies (Lane and Dunlap, 1978, Kraemer et al., 1998). This is because the initial criterion for publication and reporting (results exceeding a given statistical significance threshold) inevitably leads to the over-reporting of larger group differences that are merely a result of sampling error (which increases in smaller samples) – and these apparently larger group differences will be accompanied by correspondingly large (and inflated) effect sizes (Lane and Dunlap, 1978, Schmidt, 1992, Brand et al., 2008, Simonsohn et al., 2014).

Several statistical procedures have been suggested to quantitatively account for this inflationary bias, including Egger's test (Egger et al., 1997) and the popular Trim and Fill method (Duval and Tweedie, 2000b, a). However, more recent investigations suggest that unadjusted averages of effect size, as well as the trim-and-fill method, produce markedly inflated estimates of the true effect size (Simonsohn et al., 2014). And, unfortunately, methods such as Egger's test (Egger et al., 1997) or the *p*-curve method (Simonsohn et al., 2014) rely critically on values for the standard error, or precise *p*-values (respectively) – neither of which is routinely reported in neuroimaging studies. Therefore, we addressed this problem by employing a different method. We deflated effect sizes based on the general expected inflationary bias across study designs, as outlined in numerous theoretical and Monte Carlo simulation studies.

For instance, assume a true effect size of *d* = 0.50 (typically found in many interventions, and considered 'medium' size (Cohen, 1992)), and a significance threshold of *α* = .05 (which is the threshold employed in virtually all of the studies we examined). Under these conditions, and assuming a normal distribution of observed effects, then theoretically speaking the minimum effect size required to attain significance is ~0.62, and the average of all effect sizes that will be reported as significant is ~0.89 – nearly double the true effect size (Schmidt, 1992). From this theoretical perspective, the true effect is approximately 56% the size of the effects that have are reported because they exceed a given statistical significance threshold (Schmidt, 1992). Recent years have seen more comprehensive efforts to quantify the bias due to reporting of only significant results, and have yielded similar conclusions. One study employing 100,000 Monte Carlo simulations found that for a true effect of *d* = 0.50, the mean observed effect size (from significant results only) was 0.66 – 32% larger than the true effect (Brand et al., 2008). True effects, under these conditions, are therefore approximately 76% the size of those that are reported. Although this is considerably less inflation than suggested by others (e.g., Schmidt, 1992), these authors assumed a total sample size of $N$ = 76, which is considerably larger than samples employed in most neuroimaging experiments, including the studies investigated here (Table 2). As already noted, this inflationary bias will tend to increase with decreasing sample size (Lane and Dunlap, 1978, Kraemer et al., 1998), so the estimate of Brand et al. (2008) is likely too lenient for our purposes. In a more recent investigation, simulations showed that for a true effect size of $d$ = 0.50 the average reported effect size was around 0.82–0.85, depending on sample size and whether fixed- vs. random-effects models were employed (Simonsohn et al., 2014). This corresponds to true effects that are approximately 60% the size of those reported based only on statistically significant results.

Importantly, the size of the true effect has marked effects on this bias: the difference between true and reported effect sizes is much greater for smaller true effects, and becomes increasingly negligible when true effects are large (Simonsohn et al., 2014). For purposes of our effect size adjustments, we assumed a 'medium' effect size of $d$ = 0.50 and employed the 0.60 ratio reported by Simonsohn and colleagues (2014), which appears to us to provide the most reliable and up-to-date estimate of this potential bias. We reiterate, however, that depending on the actual size of the true effect, this bias may be greater or smaller than the 0.60 coefficient we employ here. Further, the adjustments employed here were meant mainly to ensure that the conclusions we report do not represent egregiously inflated effect sizes. Our specific deflation coefficients should be considered merely preliminary and very much open to amelioration by future empirical, theoretical, and statistical simulation research.

**Further caveats regarding effect sizes in neuroimaging.** Even given the various corrections discussed above, another concern is the interpretation of effect sizes. The interpretive guidelines laid down by Cohen (1992) were intended for the behavioral sciences, not neuroimaging, and even in the behavioral sciences questions are raised as to their utility (Rosenthal, 1996). Although it seems reasonable to use similar guidelines (e.g., the general assumption that a one-half standard deviation difference between groups is meaningful and of practical significance), rigorous discussion and elaboration of these ideas has not yet been undertaken in the field of neuroimaging. Even in the behavioral and social sciences, the interpretation of effect sizes as 'small,' 'medium,' or 'large' is ultimately arbitrary (Cohen, 1992). Moreover, the true 'practical' significance of an effect depends on many factors beyond magnitude alone. Small effects can be of large practical importance if the construct in question is highly valuable or meaningful, such as when investigating life-or-death outcomes (Prentice and Miller, 1992, Vacha-Haase and Thompson, 2004).

Despite these limitations, there have already been efforts to calculate and interpret effect sizes in meta-analyses of neuroimaging studies (e.g., (Kempton et al., 2008, Fox et al., 2014). The benefits of reporting effect size information appear to greatly outweigh the drawbacks inherent in their calculation and interpretation for neuroimaging studies. We therefore calculate and report effect sizes here to provide a general sense of the magnitude of brain activation differences reported in meditation practitioners, but we emphasize the need for caution in interpreting these results.

## Supplementary Results

*Table S1.* Studies excluded from meta-analyses (*N* = 53).

| Study | Reason for exclusion |
|---|---|
| (Khushu et al., 2000) | Original publication not locatable |
| (Ueda et al., 2000) | Original publication not locatable |
| (Bærentsen et al., 2001) | No voxel coordinates reported |
| (Kjaer et al., 2002) | Only regions-of-interest examined; dopamine release measured |
| (Ritskes et al., 2004) | No voxel coordinates reported |
| (Kakigi et al., 2005) | Single subject case study |
| (Khushu et al., 2005) | Original publication not locatable |
| (Liou et al., 2006) | No voxel coordinates reported |
| (Orme-Johnson et al., 2006) | No voxel coordinates reported |
| (Creswell et al., 2007) | No meditators involved; only 'dispositional' mindfulness scored |
| (Holzel et al., 2007) | Only regions-of-interest examined |
| (Kozasa et al., 2008) | No voxel coordinates reported |
| (Pagnoni et al., 2008) | Only regions-of-interest investigated (for contrast of interest) |
| (Beauregard et al., 2009) | Meditation experience ambiguous |
| (Baerentsen et al., 2010) | Multiple meditation types pooled in analysis |
| (Engström and Söderfeldt, 2010) | Single subject case study; no voxel coordinates reported |
| (Farb et al., 2010) | No actual meditation engaged in during scanning |
| (Goldin and Gross, 2010) | Clinical population (social anxiety disorder) |
| (Short et al., 2010) | No voxel coordinates reported |
| (Grant et al., 2011) | No actual meditation engaged in during scanning |
| (Jang et al., 2011) | No actual meditation engaged in during scanning |
| (Josipovic et al., 2011) | Only regions-of-interest examined |
| (Kilpatrick et al., 2011) | No actual meditation engaged in during scanning |
| (Kirk et al., 2011) | No actual meditation engaged in during scanning |
| (Xue et al., 2011) | No voxel coordinates reported |
| (Zeidan et al., 2011) | No voxel coordinates reported |
| (Allen et al., 2012) | No actual meditation engaged in during scanning |
| (Desbordes et al., 2012) | No actual meditation engaged in during scanning |
| (Froeliger et al., 2012a) | No voxel coordinates reported |
| (Froeliger et al., 2012b) | No actual meditation engaged in during scanning |
| (Hasenkamp and Barsalou, 2012) | No actual meditation engaged in during scanning |
| (Klimecki et al., 2012) | No actual meditation engaged in during scanning |
| (Kozasa et al., 2012) | No actual meditation engaged in during scanning |
| (Monti et al., 2012) | Clinical population (breast cancer) |
| (Pagnoni, 2012) | Only differences in skewness of fMRI signal reported |
| (Garrison et al., 2013a) | No voxel coordinates reported |
| (Garrison et al., 2013b) | No voxel coordinates reported |
| (Goldin et al., 2013) | Clinical population (social anxiety disorder) |
| (Hagerty et al., 2013) | Single subject case study |
| (Ives-Deliperi et al., 2013) | Clinical population (bipolar disorder) |
| (Klimecki et al., 2013) | No actual meditation engaged in during scanning |
| (Mascaro et al., 2013a) | No actual meditation engaged in during scanning |
| (Mascaro et al., 2013b) | No actual meditation engaged in during scanning |
| (Paul et al., 2013) | No meditators involved; only trait mindfulness scored |
| (Prakash et al., 2013) | No meditators involved; only trait mindfulness scored |
| (Taylor et al., 2013) | No actual meditation engaged in during scanning |
| (Wells et al., 2013) | Clinical population (mild cognitive impairment) |
| (Westbrook et al., 2013) | Quasi-clinical population (abstinent/craving smokers) |
| (Zeidan et al., 2013) | No voxel coordinates reported |
| (Garrison et al., 2014) | No voxel coordinates reported (for contrast of interest) |
| (Kirk et al., 2014) | No actual meditation engaged in during scanning; meditation style not reported |
| (Josipovic, 2014) | No voxel coordinates reported; preliminary data only |
| (Ding et al., 2015) | No actual meditation engaged in during scanning |

*Table S2.* Supplementary meta-analysis of focused attention meditation (excluding studies involving short-term meditation training).

| Region | Cluster Size (mm³) | Side | Peak Coordinates (x, y, z) | Peak ALE value |
|---|---|---|---|---|
| **Activations** | | | | |
| Premotor cortex | 800 | L | -36, 6, 56 (BA 6) | 0.0144 |
| Dorsal anterior cingulate cortex | 304 | M | 2, 12, 32 (BA 24) | 0.0112 |
| **Deactivations** | | | | |
| Posterior cingulate cortex | 152 | M | -6, -60, 18 (BA 30) | 0.0071 |
| Inferior parietal lobule | 144 | L | -48, -72, 30 (BA 39) | 0.0071 |

*Table S3.* Supplementary meta-analysis of mantra recitation meditation (excluding studies involving short-term meditation training).

| Region | Cluster Size (mm³) | Side | Peak Coordinates (x, y, z) | Peak ALE value |
|---|---|---|---|---|
| **Activations** | | | | |
| Premotor cortex | 896 | L | -30,4, 60 (BA 6) | 0.0137 |
| Supplementary motor area | 584 | M | 0, 12, 54 (BA 6) | 0.0111 |
| | 112 | M | -4, 2, 68 (BA 6) | 0.0080 |
| Putamen/Lateral globus pallidus | 368 | R | 28, -16, -6 | 0.0111 |
| Fusiform gyrus | 248 | R | 40, -26, -30 (BA 20/36) | 0.0093 |
| Cuneus | 160 | R | 24, -86, 26 (BA 18) | 0.0082 |
| Precuneus | 152 | L | -14, -56, 54 (BA 7) | 0.0082 |

*Table S4.* Supplementary meta-analysis of open monitoring meditation (excluding studies involving short-term meditation training).

| Region | Cluster Size (mm³) | Side | Peak Coordinates (x, y, z) | Peak ALE value |
|---|---|---|---|---|
| **Activations** | | | | |
| Temporopolar cortex | 160 | L | -45, 19, -30 (BA 38) | 0.0075 |
| Insular cortex (mid/posterior) | 160 | R | 45, -20, 13 (BA 13) | 0.0075 |
| Pre-supplementary motor area | 152 | M | 0, 8, 64 (BA 6) | 0.0080 |
| **Deactivations** | | | | |
| Thalamus (pulvinar) | 128 | R | 17, -24, 10 | 0.0084 |
| | 112 | R | 17, -30, 4 | 0.0083 |
| Parahippocampus | 104 | R | 24, -32, -9 (BA 27/35) | 0.0086 |

*Table S5.* Supplementary meta-analysis of loving-kindness/compassion meditation (excluding studies involving short-term meditation training).

| Region | Cluster Size (mm³) | Side | Peak Coordinates (x, y, z) | Peak ALE value |
|---|---|---|---|---|
| **Activations** | | | | |
| Anterior insula | 632 | R | 38, 22, 14 (BA 13) | 0.0135 |
| Parieto-occipital sulcus | 376 | R | 24, -60, 18 (BA 23/31) | 0.0115 |
| Somatosensory cortices/Inferior parietal lobule | 376 | R | 54, -26, 30 (BA 2/40) | 0.0101 |
| Postcentral gyrus/ Somatosensory cortex | 104 | L | -60, -8, 20 | 0.0083 |

**Supplementary Discussion**

*Functional neuroanatomy of focused attention meditation.*

**Activations.** If focused attention meditation so often involves strong focus on a visual stimulus or the interoceptive signals associated with respiration, why were activations not observed in visual or interoceptive cortices? Small (non-significant) clusters of activation were in fact observed in the lateral geniculate nucleus of the thalamus, which processes and relays visual information (Casagrande and Norton, 1991), as well as in the left insula, a major center for interoceptive information processing (Critchley et al., 2004, Craig, 2009, Farb et al., 2013a). These clusters can be seen in the detailed supplementary data on focused attention meditation (Fig. S1). One reason these clusters may not have been larger is due to the heterogeneity of the studies included in the meta-analysis. That is, whereas effortful cognitive control along with attention to and suppression of internal thought streams were likely common across these studies, activations related to the particular object of attentional focus may have been too disparate to attain meta-analytic significance.

**Deactivations.** The specific role of the posterior cingulate cortex in meditation has been explored more extensively in a recent study examining practitioners from diverse contemplative backgrounds (a mix of Zen, Catholic, Theravada Buddhist, and Tibetan Buddhist practitioners). Participants engaged in a focused attention meditation where the object of focus was the breath (Garrison et al., 2013a). Following an fMRI scan, participants described the subjective qualities of their meditation that corresponded to a feedback graph depicting the time course of activity from their posterior cingulate cortex throughout the scan. It was found that deactivation of the posterior cingulate cortex was associated with qualities of meditation described by terms such as 'concentration' or 'undistracted awareness' (Garrison et al., 2013a). The converse was also true: activation of posterior cingulate cortex was associated with subjective distraction and poor quality of focus during meditation (described, e.g., as 'distraction' or 'distracted awareness') (Garrison et al., 2013a). Another study from the same group reported similar findings (Garrison et al., 2013b), which are in line with our meta-analytic results.

*Functional neuroanatomy of mantra recitation meditation*

**Activations.** Activations were also observed in the fusiform gyrus (BA 20), cuneus, and the anterior precuneus (BA 7), in a location almost identical to a cluster observed in our recent meta-analysis of brain *structure* differences in experienced meditation practitioners (see Fig. 2g in (Fox et al., 2014). Both clusters were observed in a putative 'anterior' subregion of the precuneus with strong connections to somatomotor cortices and insula, which may play a role in coordination of motor behavior as well as the generation of motor imagery (Hanakawa et al., 2003, Malouin et al., 2003, Cavanna and Trimble, 2006). A speculative role for anterior precuneus, then, may be in assisting in the overt or mental repetition of the mantra. Alternatively, or in addition to this interpretation, the activation of the precuneus as well as the fusiform gyrus and cuneus may indicate that practitioners evoke visual imagery along with mantra repetition. Detailed subjective reports will be necessary to more fully elucidate the processes activated during mantra recitation, and by extension the meaning of these activations.

*Functional neuroanatomy of open monitoring meditation.*

**Deactivations.** Many other smaller (non-significant) clusters of deactivation were observed. One deactivation that accords well with the goal of open monitoring meditation was located in the rostromedial prefrontal cortex (medial BA 10) – as opposed to the cluster of *activation* observed in lateral BA 10 (see Fig. S3). A central goal of open monitoring is to dampen the tendency to become ensnared in one's thoughts and emotions, and to not turn isolated mental content into extended streams of self-referential thinking (Goenka, 2000). As medial prefrontal cortex has been so consistently implicated in these kinds of mind-wandering (Fox et al., 2015) and self-referential thought (Northoff et al., 2006), a simple explanation is that its deactivation represents the successful execution of open monitoring (Farb et al., 2007).

Consistent with this explanation are other small (non-significant) clusters of deactivation observed bilaterally in the subgenual anterior cingulate cortex (or ventromedial prefrontal cortex; BA 25) and bilaterally in the medial temporal lobe (BAs 27, 28, and 36, including the hippocampus) – both of which are similarly implicated in the default mode network (Buckner et al., 2008) as well as mind-wandering and related spontaneous thought processes (Fox et al., 2015). Also consistent with such an interpretation were small clusters of deactivation bilaterally in the posterior cingulate cortex (BAs 23, 29, 30, and 31), likewise involved in the default mode network (Buckner et al., 2008) and mind-wandering (Fox et al., 2015). All of these regions of course play many other roles, but nonetheless the widespread deactivation of so many important default network hubs, which are consistently activated during mind-wandering (Fox et al., 2015), is suggestive of a lowered frequency of, and/or a decreased involvement in, one's thoughts.

Various other small clusters of deactivation were also observed, for instance in the left lingual gyrus of the occipital lobe (BA 18/19). We have previously found the lingual gyrus to be consistently recruited during both nighttime dreaming (Fox et al., 2013) and waking daydreaming/mind-wandering (Domhoff and Fox, 2015, Fox et al., 2015), and lesions to this region eliminate all visual imagery in dreams, while sparing the dream experience itself (Solms, 1997). In line with our suggestions regarding deactivation of default network areas, it may be that open monitoring dampens the visual imagery so characteristic of mind-wandering during normal waking (Fox et al., 2013), as well as during formal meditation practice (Austin, 1999).

*Functional neuroanatomy of loving-kindness and compassion meditation*

**Activations.** A notable but non-significant cluster of activation was observed in the anterior cingulate cortex (BA 24), very near to a meta-analytic cluster reported in a meta-analysis of empathy for the pain of others (see Fig. 2 in (Lamm et al., 2011)). This cluster was located in a more anterior location than those identified in the other types of meditation. As with somatosensory cortices and inferior parietal lobule (discussed above), this area is also consistently implicated in pain perception, particularly its affective (as opposed to sensory) component (Grant et al., 2011). Similarly, then, this area might play overlapping roles in imagining the pain and distress of others, as well as empathizing with their pain and distress.

**Deactivations.** No significant deactivations were observed, and very few non-significant deactivations were apparent (Fig. S4). As noted above, however, our meta-analysis of loving-kindness and compassion meditations was based on the fewest experiments and smallest number of foci (Table 1), and therefore had lower statistical power than the other meta-analyses. The relative lack of activations and deactivations in this form of meditation practice more likely reflects this meta-analytic limitation rather than any inherent feature of the practice itself. More research is therefore needed to better illuminate the activations and deactivations associated with this group of meditation practices.

*Convergent findings across meditation categories*

**Insula**. The relatively stronger (and statistically significant) recruitment of the insula in open monitoring and loving-kindness/compassion meditation is further consistent with the character of these practices. These practices tend to focus heavily on awareness of the body and emotions (open monitoring), or empathy, mentalizing, and perspective-taking (loving-kindness/compassion). For open monitoring, there is growing evidence that practitioners who heavily emphasize this practice show heightened interoceptive and exteroceptive bodily awareness. Support for this enhanced awareness comes from multiple methodologies, including behavioral (Fox et al., 2012), physiological (Sze et al., 2010), magnetoencephalographic (Kerr et al., 2013), and fMRI (Farb et al., 2007, Farb et al., 2013b) research.

Beyond body awareness and interoception, the insula is also reliably implicated in empathy and the perspective-taking or mentalizing that empathy often entails (Lamm et al., 2011, Bernhardt and Singer, 2012). The significant insula cluster observed in loving-kindness and compassion meditation, then, suggests an alternate function that the insula might serve in meditations involving empathy, imaging the pain and suffering of others, and the cultivation of compassionate and prosocial responses to this distress.

Although further speculation about the insula's particular function(s) across various types of meditation seems unwarranted at this early stage, both morphometric meta-analysis (Fox et al., 2014) and the present functional meta-analysis agree that it plays an important role across several forms of meditation practice.

*Are contemplative practices dissociable by electrophysiological and neurochemical measures?*

One group has attempted to integrate EEG findings from across many meditation styles, and similar to our claims here, argued that various forms of meditation are preferentially associated with differing electrical rhythms in the brain (e.g., theta, beta, gamma, and so on) (Travis and Shear, 2010). Others, however, have suggested major commonalities in EEG signatures across many different meditation styles (Lehmann et al., 2012). Whether these findings can be integrated with the current arguments from blood-flow or blood-oxygenation-dependent measures remains an important question for future research and meta-analytic synthesis.

The ability of positron emission tomography (PET) scanners to investigate the activity of particular neurotransmitters and neuromodulators has hardly been exploited in meditation research. A single study examining *yoga nidra* using PET has reported increased dopamine release during this form of practice (Kjaer et al., 2002), but to our knowledge no research has followed up on these initial findings. Various meditation practices consistently lead to subjective changes in relaxation, levels of alertness and vigilance, engagement in visionary and dream-like experiences, and so on – all of which are known to be heavily driven by changes in levels of various neurotransmitters and neuromodulators (Perry et al., 2002). As such, there is good reason to believe that both the tonic and phasic activity profiles of various neurotransmitters might vary markedly across different practice categories. Given the paucity of research to date, however, these remain open questions to be addressed by future research.

*Other methodological challenges in the neuroimaging of meditation*

**'Overshadowing.'** As discussed in the Introduction, we assumed a *one-to-many* logical relationship (Coombs et al., 1970, Cacioppo and Tassinary, 1990) between a given psychological meditation state ($\Psi$) and its diverse neural correlates ($\Phi$): one form of meditation practice (e.g., focused attention) would be accompanied by numerous (many) but hypothetically consistent neurophysiological responses. If a given meditation state is strongly colored by a practitioner's contemplative background (experience with a variety of other meditation techniques), however, then this assumption no longer holds, and the relationship assumes a *many-to-many* form, making it much more difficult to draw any reliable conclusions.

Unfortunately, it seems likely that this is the situation in at least some of our sample: as already noted (Table 1), some investigators had the same sample of practitioners engage in multiple meditation techniques in the same study. This was possible because many practitioners are familiar with, and have long practiced, multiple techniques. Clearly, the particular ratio of experience with different techniques will vary among individual practitioners, and/or practitioners from differing meditation traditions, adding an additional (and unknown) degree of heterogeneity. For instance, some traditions focus heavily on mantra meditation practice (e.g., Transcendental Meditation), although their practitioners may of course also be familiar with other practices. Other traditions (e.g., Theravada Buddhist) tend to focus more on open monitoring techniques. The central point is that these various emphases are largely unknown. When a practitioner participates in a neuroimaging study and is asked to engage in a particular practice for a short period of time (usually only a few minutes) in a brain scanner, their traditional contemplative background and the emphasis they have placed on various techniques over their years of practice may greatly influence their manner of meditation practice (and the concomitant neural correlates investigators record and report) – but these diverse contemplative backgrounds are rarely investigated or reported (typically, total hours or years of meditation experience is the only data collected). Although total hours of experience is a useful preliminary measure, the incredible diversity of meditation techniques clearly calls for more detailed measures in future work.

**Integration of functional neuroimaging methods with behavioral measures.** In our previous meta-analysis, we noted that numerous studies attempted to directly link brain structure changes to behavior change or scores on various self-report measures, but almost all of the reported correlations were statistically marginal and/or statistically non-independent. Among the present pool of studies, a few have reported modest correlations between meditation expertise (i.e., total hours of practice) and differences in brain activation during a given meditation state (e.g., (Lazar et al., 2000, Brefczynski-Lewis et al., 2007, Farb et al., 2013b). Few, however, have related brain activation differences *during meditation states* to actual behavior, or even to self-report questionnaires (e.g., for anxiety or mindfulness): of the 26 studies included here, only four correlated their findings with behavioral outcomes other than meditation experience or expertise. We acknowledge that this is an extremely challenging problem: of innumerable potential outcomes, what behavior should be measured? What self-report questionnaires should be considered the most relevant and reliable? Can reliable correlations be obtained with the small samples typical of neuroimaging studies? Is activation during the meditation *state* even the relevant independent variable – i.e., should these correlations instead be sought using baseline, *trait* levels of brain activity, outside of the meditation practice itself (see 5.12.4)?

As noted above, however, several of the more recent fMRI studies *have* attempted to integrate neuroimaging findings with behavioral measures other than just meditation experience itself (Lee et al., 2012, Lutz et al., 2013, Weng et al., 2013, Lutz et al., 2014). For instance, Weng et al. (2013), in a study of compassion meditation, reported that those with greater brain activation in dorsolateral prefrontal cortex and inferior parietal cortex during the meditation were more likely to redistribute funds more equitably in an economic game they subsequently participated in. This study therefore demonstrated not only that compassion meditation training leads to differential neural responses to the suffering of others and to increased altruistic behaviors in an economic game – it also linked these findings together, providing evidence that specific contemplative training leads to changes in brain activity changes which in turn are related to meaningful behavioral outcomes. This multifaceted approach remains rare, but provides a model for future research, which can (and should) integrate neuroimaging measures with behavioral and first-person report outcomes.

Further, several of the studies excluded from our meta-analysis that examined 'trait' differences in meditators (Table S1) or that employed meditation in clinical populations have also assessed various outcome measures as correlated with their neuroimaging results (Table S1). However, we hasten to reiterate the caveats we have raised before regarding such correlational analyses (cf. Fox et al., 2014): for instance, correlation coefficients are highly unstable with small sample sizes (Glass and Hopkins, 1970), and efforts to correlate neuroimaging findings with behavioral measures are highly susceptible to 'circular' or non-independent analysis, which can grossly inflate correlation coefficients (Vul et al., 2009, Hupé, 2015). Nonetheless, in order for the neuroimaging of meditation to move beyond cartography of brain activation differences and toward an integrative science linking psychological training with brain and behavior, these methodological and statistical difficulties will need to be overcome.

## Supplemental References

*Figure S1.* Activations and deactivations associated with focused attention mediation.

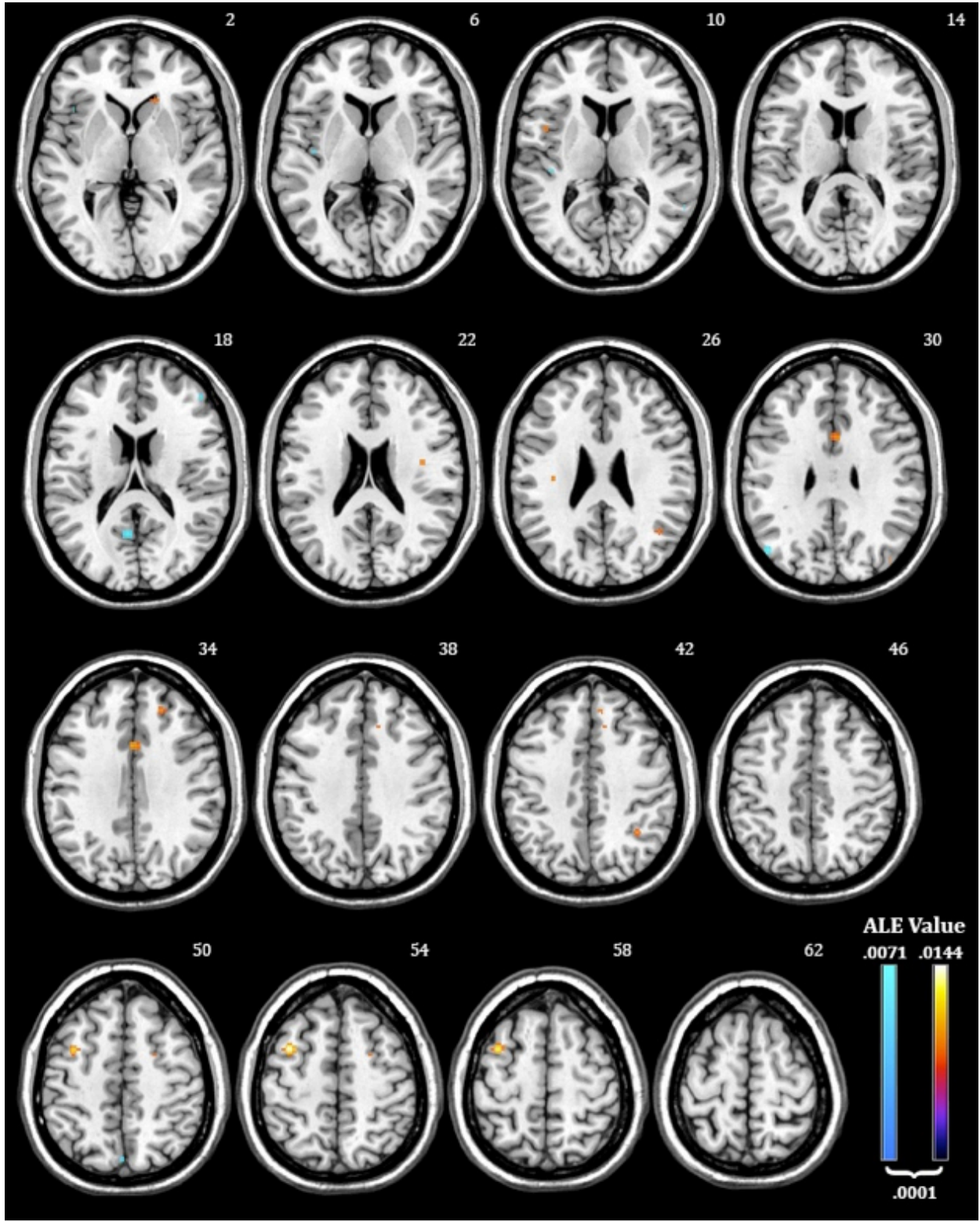

*Figure S2.* Activations and deactivations associated with mantra recitation mediation.

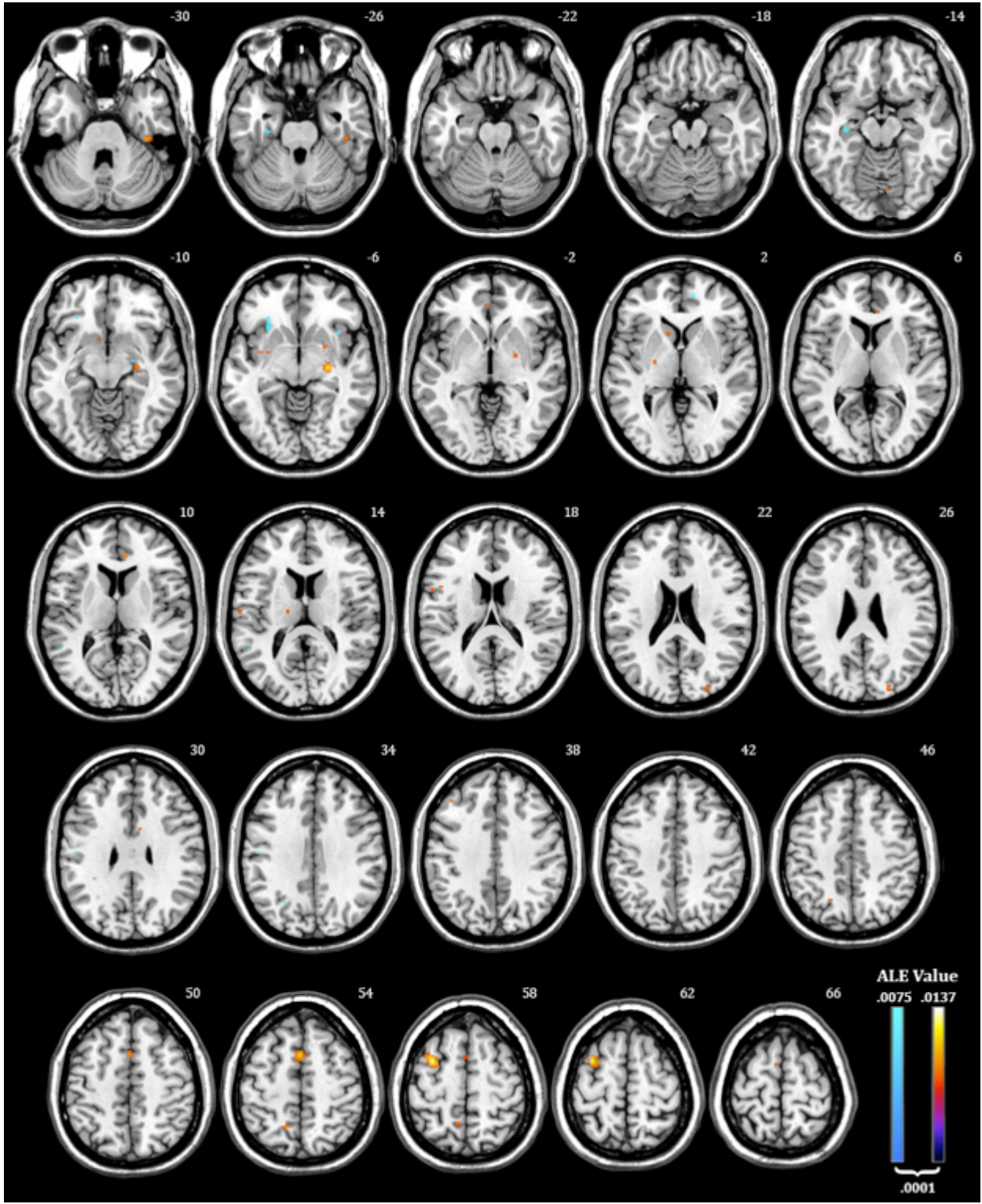

*Figure S3.* Activations and deactivations associated with open monitoring mediation.

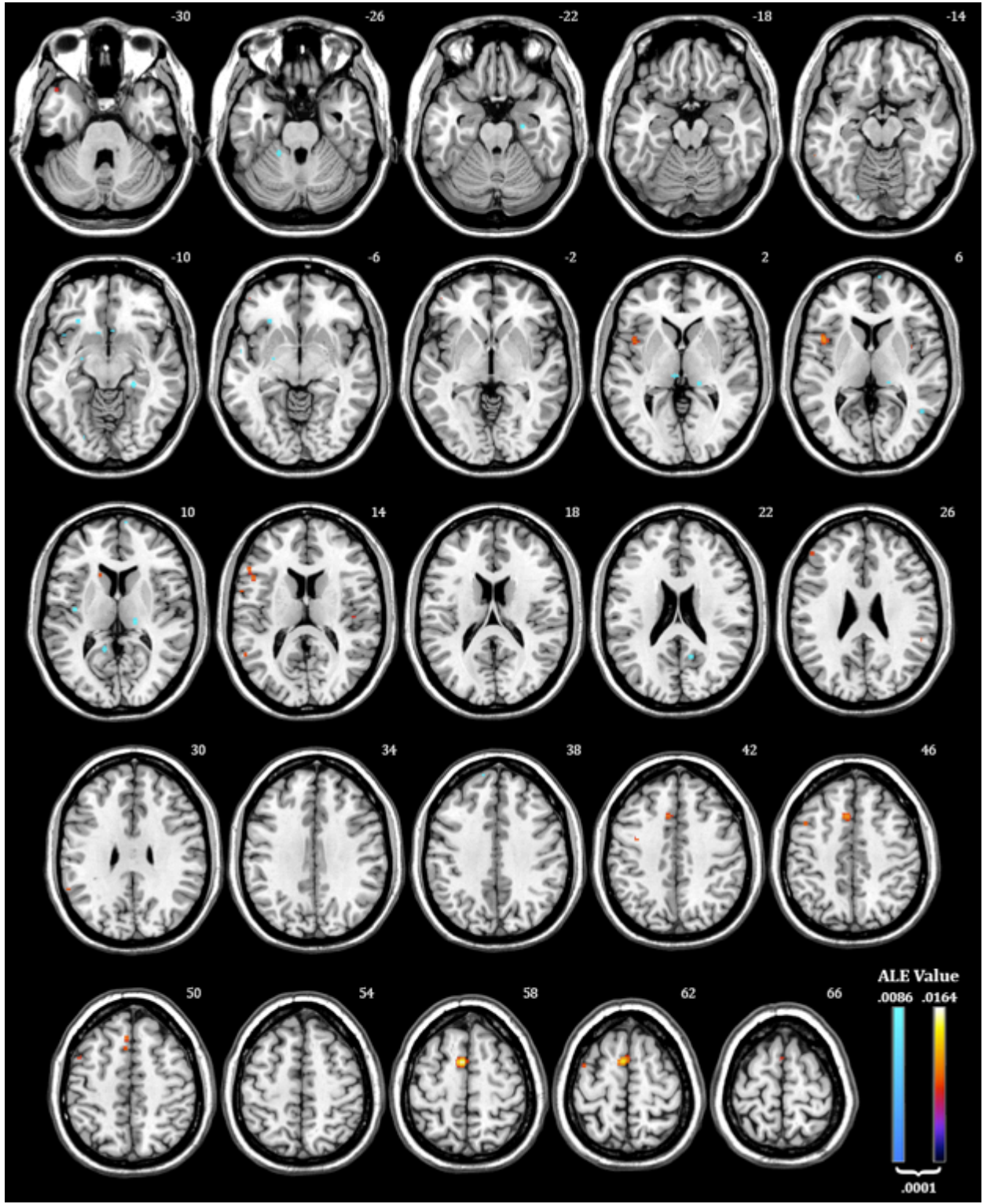

*Figure S4.* Activations associated with loving-kindness and compassion mediation.

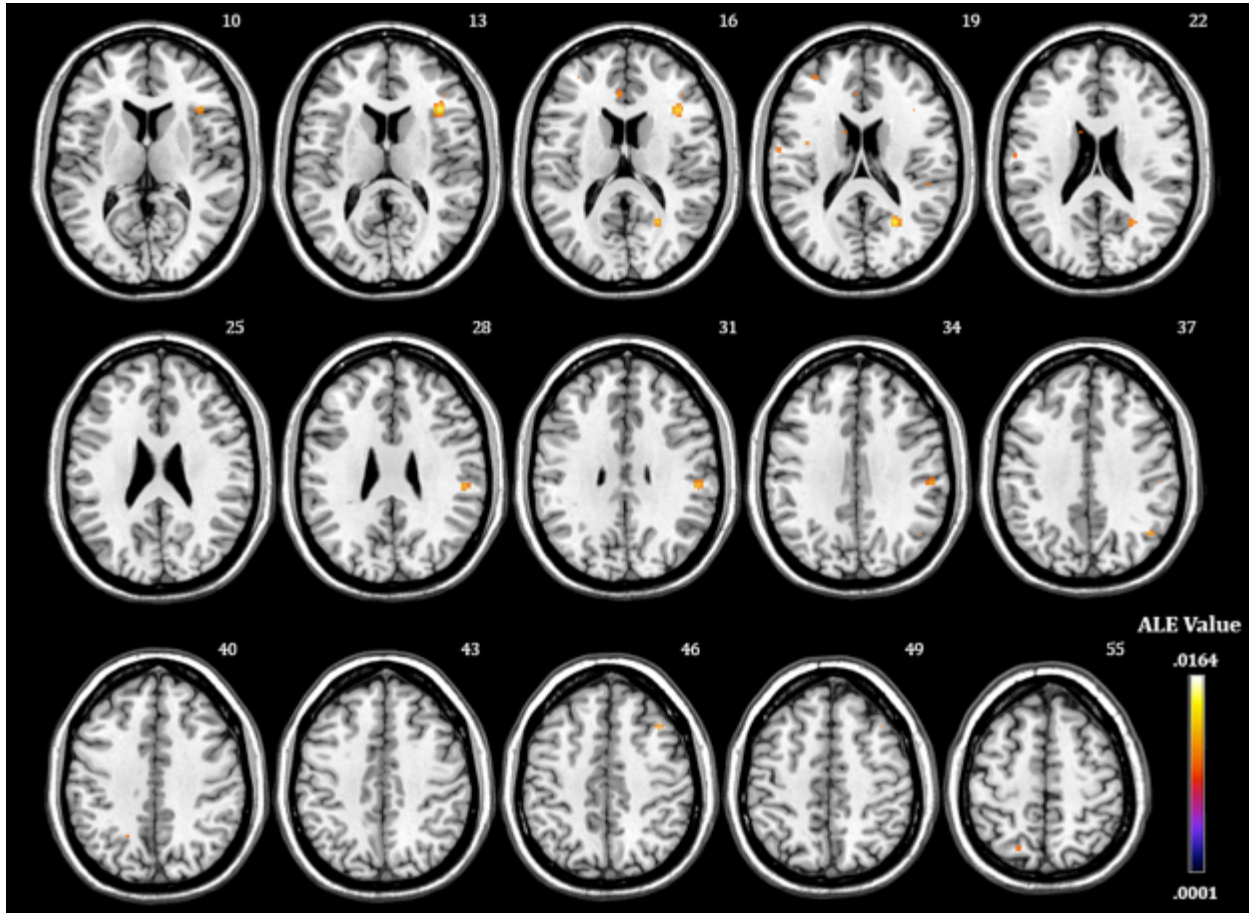